\documentclass[12pt,a4paper]{article}

\usepackage{cancel}

\usepackage[english]{babel}
\usepackage[autostyle,english=british]{csquotes}

\usepackage[backend=bibtex, citestyle=numeric-comp, bibstyle=ieee]{biblatex}
\usepackage{mathtools,amssymb,bbm}
\usepackage{braket,tensor}
\usepackage{xcolor}
\usepackage{textgreek}
\usepackage{nicefrac}
\usepackage{amssymb}
\usepackage{amsfonts}
\usepackage[normalem]{ulem}

\usepackage{enumitem}
\usepackage[a4paper, total={6in, 9in}]{geometry}
\usepackage[hidelinks]{hyperref}
\definecolor{jade}{RGB}{0, 168, 107}
\hypersetup{
 colorlinks,
 linkcolor={jade},
 citecolor={jade},
 urlcolor={jade}
}

\renewcommand{\theequation}{\thesubsection.\arabic{equation}} \csname
@addtoreset\endcsname{equation}{subsection}

\newcommand{\diff}{\mathrm{d}}
\newcommand{\de}{\partial}

\newcommand{\vtheta}{\vartheta}
\newcommand{\vphi}{\varphi}
\newcommand{\vrho}{\varrho}

\makeatletter
\newenvironment{equations}[1][]{\subequations\ifx\relax#1\relax\else\label{#1}\fi\align\ignorespaces}{\endalign\ignorespacesafterend\endsubequations}
\def\@spliteq#1{\begin{equation}\begin{split}#1\end{split}\end{equation}}
\def\splitequation{\collect@body\@spliteq}

\makeatother

\makeatletter
\newcommand{\firstsectioneqnums}{%
  \renewcommand{\theequation}{\thesection.\arabic{equation}}%
  \@addtoreset{equation}{section}%
}
\newcommand{\restoreeqnums}{%
  \renewcommand{\theequation}{\thesubsection.\arabic{equation}}%
  \@addtoreset{equation}{subsection}%
}
\makeatother

\begin{document}

\begin{titlepage}

\begin{center}
{\LARGE \bf Worldline Formulations of\\[2mm]Covariant Fracton Theories}
\vskip 1.2cm
Filippo Fecit$^{\,a,b}$ and Davide Rovere$^{\,c,d}$
\vskip 1cm
$^a${\em Dipartimento di Fisica e Astronomia ``Augusto Righi", Universit{\`a} di Bologna,\\
via Irnerio 46, I-40126 Bologna, Italy}\\[2mm]
$^b${\em INFN, Sezione di Bologna, Bologna, Italy}\\[2mm]
$^c${\em Dipartimento di Fisica e Astronomia ``Galileo Galilei'', Universit\`a di Padova,\\
via Marzolo 8, 35131 Padua, Italy}\\[2mm]
$^d${\em INFN, Sezione di Padova, Padua, Italy}
\end{center}
\vskip 1cm

\abstract{We develop worldline formulations of covariant fracton gauge theories. These are a one-parameter family of gauge theories of a rank-two symmetric tensor field, invariant under a scalar gauge transformation involving a double derivative. These theories, which can be interpreted as linearized gravity theories invariant under longitudinal diffeomorphisms, provide a covariant framework for studying Lorentz-breaking fracton quasiparticles, which are excitations with restricted mobility due to dipole-moment conservation.

We construct three worldline models. The first two are obtained by deducing their constraint structure directly from the spacetime gauge transformations. By applying BRST quantization, we show that these models reproduce the BV spectrum and the associated BRST transformations of two specific fracton theories. The third model is defined as a deformation of the second one: although free, it is analyzed by drawing inspiration from the standard treatment of interacting worldline systems, and is shown to capture almost the entire family of covariant fracton theories.

Finally, we discuss the gauge-fixing, comparing the BV-BRST spacetime perspective with the worldline analogue of the ``Siegel gauge" employed in string field theory.}

\end{titlepage}

\tableofcontents

\section{Introduction}

\firstsectioneqnums  
The aim of this paper is to investigate possible worldline formulations of covariant fracton theories, continuing the broader program of reproducing the Batalin-Vilkovisky (BV) formulation of spacetime field theories \cite{Batalin:1981jr, Batalin:1983ggl, Barnich:2000zw} from a first-quantized perspective.

Covariant fracton gauge theories, introduced for the first time in \cite{Blasi:2022mbl}, are a family of gauge theories of a rank-two symmetric tensor $h_{\mu\nu}$ with gauge symmetry given by the double derivative of a scalar parameter $\Lambda$
\begin{equation}\label{GaugeIntro}
\delta_\Lambda\,h_{\mu\nu} = \de_\mu\,\de_\nu\,\Lambda.
\end{equation} 
The quadratic action, left invariant by this gauge transformation, can be written as
\begin{equation}\label{FractonActionWithF}
S^{\text{(fr)}} = \int\diff^D x\,(\alpha\,f^{\mu\nu\vrho}\,f_{\mu\nu\vrho} + \beta\,f^{\mu\nu}{}_\nu\,f_{\mu\vrho}{}^\vrho),
\end{equation}
where $\alpha$ and $\beta$ are free constant parameters\footnote{Note that covariant fracton theories actually depend only on one free parameter, as either $\alpha$ or $\beta$ can be absorbed by a field redefinition. We will keep the two constants separate for future convenience.} and the theory is defined in a $D$-dimensional Minkowski spacetime. $f_{\mu\nu\vrho}$ is the gauge invariant field strength of $h_{\mu\nu}$, defined by
\begin{equation}\label{ActionIntro}
f_{\mu\nu\vrho} = \de_\mu\,h_{\nu\vrho} - \de_\nu\,h_{\mu\vrho},
\end{equation}
such that
\begin{equation}
\delta_\Lambda\,f_{\mu\nu\vrho} = 0.
\end{equation}

In recent years, exotic excitations with restricted mobility, called ``fractons", have been extensively studied, originally in the context of condensed matter physics \cite{Pretko:2017xar, Burnell:2021reh, Makino:2025mzo}. Fracton excitations appear in lattice spin models \cite{Haah:2011drr, Vijay:2015mka}, whose low-energy continuous limit \cite{Affleck:1986} can be captured by effective field theories \cite{Pretko:2020cko, Seiberg:2020bhn}. These models are characterized by single charged particles and dipoles with not only conserved charge $Q = \int\diff^{D-1}x\,\vrho$, but also conserved dipole moment $\vec{P}=\int\diff^{D-1}x\,\vec{x}\,\vrho$, that is,  $\frac{\diff}{\diff t}\,Q = 0 = \frac{\diff}{\diff t}\,\vec{P}$. Namely, the dipole-moment conservation implies for a single charged particle to be fixed in space, whereas dipoles are free to move. 

These features are reproduced by a pair of gauge fields, $\vphi$ and $A_{ij}$, where $i,j$ are space indices, with gauge transformations 
\begin{equation}\label{GaugeNonLorentz}
\delta \vphi = \de_t\,\Lambda, \quad \delta A_{ij} = \de_i\,\de_j\,\Lambda,
\end{equation}
with $\Lambda$ a scalar parameter. Indeed, a coupling of the form $-\vrho\,\vphi + J^{ij}\,A_{ij}$, for some fractonic matter current $\vrho$ and $J^{ij}$, leads to a higher-derivative continuity equation $\de_t\,\vrho + \de_i\,\de_j\,J^{ij} = 0$, which implies the conservation of both $Q$ and $\vec{P}$.

A concrete example of fractonic matter theory is studied in \cite{Pretko:2018jbi} (see also \cite{Pretko:2020cko}), where the theory of a complex scalar field $\Phi$ is discussed. Such theory, besides the $U(1)$ invariance $\delta\Phi = i\,\alpha\,\Phi$, responsible of the charge conservation, enjoys also a dipole symmetry $\delta\Phi = i\,\beta_i\,x^i\,\Phi$, for some constant vector $\beta_i$, which implies the dipole-moment conservation. The gauging of such symmetries, as discussed in \cite{Bidussi:2021nmp}, requires the aforementioned pair of gauge fields $\vphi$ and $A_{ij}$, with gauge parameter $\Lambda = \alpha(x) + \beta_i(x)\,x^i$.

Fractonic matter theories and fractonic gauge theories manifestly break the Lorentz invariance, since the gauge transformations \eqref{GaugeNonLorentz} involve a different number of space and time derivatives. Nevertheless, the transformations \eqref{GaugeNonLorentz} can be embedded in the covariant gauge transformation \eqref{GaugeIntro} of $h_{\mu\nu}$, by identifying $h_{0\mu} = \de_\mu\,\vphi$, $h_{ij}=A_{ij}$ \cite{Bertolini:2022ijb}. The covariant minimal coupling $j^{\mu\nu}\,h_{\mu\nu}$, for some covariant fractonic current $j^{\mu\nu}$, which, as a consequence of gauge invariance, satisfies the equation $\de_\mu\,\de_\nu\,j^{\mu\nu} = 0$, becomes the original continuity equation, if $\vrho = \de_t\,j^{00} + \de_i\,j^{0i}$ and $J^{ij} = j^{ij}$. Therefore, the covariant theory provides a larger framework in which the genuine fracton models are embedded. 

But since the distinctive feature of fractonic excitations -- the reduced mobility -- is by definition Lorentz-breaking, the label ``fracton" for the covariant gauge theory defined by the action \eqref{FractonActionWithF} is misleading. Nevertheless, we maintain it, in accordance with existing literature on the topic (\cite{Blasi:2022mbl} and the subsequent works). Moreover, we stress that by ``covariant fracton gauge theories" we mean the pure Lorentz-invariant models of the gauge field $h_{\mu\nu}$, with action \eqref{FractonActionWithF}, without specifying any matter theory coupled to it.\footnote{The Lorentz-breaking fractonic matter theory discussed in \cite{Pretko:2018jbi} can be covariantized in a natural way, as done in \cite{Afxonidis:2023pdq}, thereby loosing the restriction in the mobility of the particle excitations, which is the distinctive feature of ``true" fractons.}

Covariant fracton gauge theories are interesting \emph{per se}, covering a wider scope than the reasons why they were originally introduced. Indeed, on the one hand, they can be seen as a higher rank electrodynamics \cite{Bertolini:2022ijb}, because of the scalar gauge invariance \eqref{GaugeIntro} and the field strength \eqref{ActionIntro}, involving a single derivative. On the other hand, covariant fracton gauge theories can be thought of as a family of generalizations of linearized gravity \cite{Blasi:2022mbl, Bertolini:2023juh}, where the invariance under linearized diffeomorphisms in the latter theory is restricted to longitudinal diffeomorphisms in the former. While the gauge invariance of linearized gravity reduces the propagating degrees of freedom to spin-2 polarizations only, in covariant fracton gauge theories propagating spin-1 and spin-0 excitations remain \cite{Afxonidis:2023pdq}, which lead to instabilities \cite{Afxonidis:2024tph}. They could be cured by considering suitable interactions -- a first attempt is presented in \cite{Bertolini:2024apg}. The connection with gravity theories in the torsion formulation, as sketched in \cite{Rovere:2025nfj}, may help clarify this issue.

An interesting perspective has been recently suggested in \cite{Hinterbichler:2025ost}. Exploiting the similarities of the gauge transformation \eqref{GaugeIntro} with that of a partially massless spin-2 field in de Sitter spaces,  relativistic fractonic matter is coupled to the gauge field, and the Higgs mechanism for the partially massless field is studied. As a result, a superconducting phase is induced, characterized by a condensation of fractonic matter.

Further progress in understanding covariant fracton theories may benefit from exploring them through a different perspective in the context of worldline formalism \cite{Schubert:2001he, Edwards:2019eby, Newbook:2025}. This approach offers an alternative to conventional second-quantized methods. On the one hand, it enables more efficient perturbative computations, as illustrated by recent calculations of heat-kernel coefficients and scattering amplitudes \cite{Bastianelli:2023oca, Fecit:2024jcv, Ahumada:2025dyc, Bastianelli:2025xne}. On the other hand, being intrinsically a functional method, it is well-suited for non-perturbative analyses: in particular, it has proven useful in the study of strong-field phenomena, including the non-perturbative production of particle-antiparticle pairs \cite{Fecit:2025kqb, Bastianelli:2025khx, Ahumada:2025vex, Ilderton:2025umd, Semren:2025dix}. 

Our goal is to construct first-quantized models capable of reproducing covariant fracton theories as spacetime theories. We will employ BRST quantization, which naturally yields the complete spectrum of spacetime fields required by the BV formalism, up to local redefinitions of the fields. Let us note that worldline-based approaches to (Lorentz-breaking) fractons models have appeared in the literature \cite{Casalbuoni:2021fel, Bidussi:2021nmp}. Instead, our aim is to describe the covariant theory of the gauge field $h_{\mu\nu}$, defined by \eqref{ActionIntro}, from a wordline perspective, in a manifestly covariant way.

The worldline program for field theories describing spin-$s$ particles has been known for a long time. The so-called $O(\mathcal{N})$ spinning particle models provide first-quantized descriptions of massless and massive spin-$s$ particles propagating on flat spacetime \cite{Berezin:1976eg, Barducci:1976qu, Brink:1976sz, Gershun:1979fb, Howe:1989vn, Bastianelli:2005uy, Bastianelli:2005vk}. Crucially, these models employ fermionic variables to encode spin degrees of freedom and rely on local $\mathcal{N}=2\,s$ worldline supersymmetries to ensure unitarity. An alternative approach describes the spin degrees of freedom using bosonic variables \cite{Bengtsson:2004cd, Hallowell:2007qk, Bastianelli:2009eh}, a formulation that has also seen renewed interest in recent years \cite{Bonezzi:2024emt, Bastianelli:2025khx}. In this work, we will adopt a similar framework based on bosonic phase-space variables.

To describe more general gauge theories, the BRST formalism offers the most promising path forward \cite{Henneaux:1994lbw}. This technique actually finds its most successful application in the consistent coupling of worldline models to background fields, as illustrated in the seminal work \cite{Dai:2008bh}. There it was observed that a fully nilpotent BRST charge is necessary only on a restricted subspace of the extended Hilbert space. This relaxed condition allows for consistent interactions and has been applied successfully to couple massless and massive spin-2 particles to gravitational backgrounds \cite{Bonezzi:2018box, Bonezzi:2020jjq, Fecit:2023kah} and massless/massive spin-1 particles to non-abelian/abelian backgrounds respectively \cite{Bonezzi:2024emt, Bastianelli:2025khx}. 
Relevant to this work, BRST methods have proven essential for the worldline formulation of a broader class of spacetime field theories. Notably, topological theories such as Chern-Simons theory have been formulated as worldline models \cite{Witten:1992fb}, and even supersymmetric theories such as linearized $D=10$ super Yang-Mills and $D=11$ supergravity have been described using a first-quantized language \cite{Berkovits:2001rb, Berkovits:2019szu}.

To construct worldline models for covariant fracton theories, we follow the general strategy outlined in \cite{Berkovits:2001rb}. We begin with a phase space consisting of standard coordinates and momenta $(x^\mu,p_\mu)$, augmented by suitable bosonic oscillator variables. In contrast to spinning particle models, where these oscillators are introduced to describe spin, their role is motivated here by the structure of the gauge transformation \eqref{GaugeIntro}, which naturally suggests the introduction of constraints proportional to $\sim p_\mu\, p_\nu$. The additional variables then serve to saturate the momenta.

We find multiple possible worldline formulations that realize this structure. The first, which we refer to as ``tensor model", introduces a pair of symmetric tensor variables $(\alpha^{\mu\nu}, \bar{\alpha}_{\mu\nu})$, complex conjugate to each other, to saturate the double-momentum constraint. By imposing the constraint $L=\alpha^{\mu\nu}\,p_\mu \,p_\nu$, along with its conjugate $\bar L$, we find that their Poisson brackets close quadratically into a standard Hamiltonian constraint, $H=p^2$ for a free theory, forming a first-class algebra of constraints \cite{VanHolten:2001nj}. In this sense, the model shares many features with conventional particle models. Proceeding with BRST quantization, we establish that it successfully describes the covariant fracton gauge theory characterized by
\begin{equation}
\text{Tensor model} \rightarrow 2\,\alpha - \beta = 0, 
\end{equation}
in all spacetime dimensions $D$.

An alternative approach, which we call ``vector model", employs a pair of vector oscillators $(\alpha^\mu, \bar{\alpha}_\mu)$. This leads us to implement the constraint $L=\alpha^\mu\,p_\mu\,\alpha^\nu\,p_\nu$ this time, together with $\bar L$, which are quadratic not only in the momenta but also in the bosonic oscillators. As a consequence, the algebra of constraints becomes more involved. Nevertheless, it is possible to identify a suitable third constraint $\ell$ such that the set of constraints forms a first-class algebra with structure functions. Although it would in principle be possible to implement a standard Hamiltonian constraint as well, we find that its inclusion fails to yield a consistent worldline formulation for covariant fracton theories. We therefore choose to leave the model without a Hamiltonian constraint, and nevertheless find that we can  successfully reproduce again a specific covariant fracton theory without encountering any apparent consistency issue. This may seem surprising, since in conventional particle models the Hamiltonian constraint plays a pivotal role; we will comment on this point in due course. The vector model turns out to describe the covariant fracton gauge theory with the following parameters:
\begin{equation}
\text{Vector model} \rightarrow 2\,\alpha + 3\,\beta = 0,
\end{equation} 
in all spacetime dimensions $D\neq 4$.

In order to describe a broader class of fracton models, we then consider a deformation of the vector model, which we refer to as ``deformed vector model". Its treatment draws inspiration from techniques usually employed in interacting worldline models, which prove highly effective for our purposes. We find that this third, deformed model is capable of reproducing all known fracton theories, except three particular cases:
\begin{equation}
\text{Deformed vector model} \rightarrow 
\text{any}\;\alpha,\beta,\;\text{s.t.}\;
2\,\alpha - \beta \neq 0, \; \beta \neq 0,\;
2\,\alpha + (D-1)\,\beta \neq 0.
\end{equation}
The first excluded case is already captured by the tensor model. The last case is the traceless limit, in which the gauge symmetry of the theory is larger, since it also includes the invariance under local scaling.

\bigskip

The paper is organized as follows. In Section~\ref{spacetime}, we review the key features of covariant fracton theories from a spacetime perspective, studying the corresponding BV formulation. In Section~\ref{worldline} the tensor model (Section~\ref{modelI}), the vector model (Section~\ref{modelII}), and the deformed vector model (Section~\ref{modelIII}) are introduced and discussed in details. In Section~\ref{gauge-fixing}, the gauge-fixing procedure is discussed, both from the spacetime perspective and from the worldline one, and the two approaches are compared. In Section~\ref{conclusions} a summary of the results and an outline of possible directions for future developments are discussed. Appendix~\ref{AppendixConvNot} collects our notations and conventions, namely those regarding BV formalism, fermionic variables and inner product in graded Hilbert spaces.

\section{BV formulation of covariant fracton theories} \label{spacetime}

Covariant fracton theories are defined as the gauge theories of the symmetric rank two tensor $h_{\mu\nu}$ with the gauge transformation 
\begin{equation} \label{gauge}
\delta_\Lambda\,h_{\mu\nu} = \de_\mu\,\de_\nu\,\Lambda.
\end{equation}
The most general quadratic action, invariant under the previous gauge transformation, reads\footnote{The relation with the constants using in \cite{Blasi:2022mbl} is $\alpha =  (a-1)/2$, $\beta = 1$; in \cite{Bertolini:2023juh} and in \cite{Afxonidis:2024tph} is $\alpha = (g_2-g_1)/2$ and $\beta = g_1$.}
\begin{align}
S^{\text{(fr)}} &=\int\diff^D x\,(\alpha\,f^{\mu\nu\vrho}\,f_{\mu\nu\vrho} +\beta\,f^{\mu\nu}{}_\nu\,f_{\mu\vrho}{}^\vrho)  \label{actionF}\\
& = \int\diff^D x\,(-2\,\alpha\,h^{\mu\nu}\,\de^2\,h_{\mu\nu} -\beta\,h\,\de^2\,h- (2\,\alpha-\beta)\,\de_\mu\,h^{\mu\nu}\,\de^\vrho\,h_{\vrho\nu} + 2\,\alpha\,h\,\de_\mu\,\de_\nu\,h^{\mu\nu}),\label{action}
\end{align}
where $h=h_\mu{}^\mu$, and $f_{\mu\nu\varrho}$ is the gauge invariant field strength of $h_{\mu\nu}$, defined by\footnote{The gauge invariant field strength $f_{\mu\nu\vrho}$ was used in \cite{Rovere:2024nwc} in studying the BRST cohomology of covariant fracton theories and in \cite{Rovere:2025nfj} in writing the action as in \eqref{actionF} and comparing with the linearization of the torsion formulation of General Relativity and its extensions. As pointed out in \cite{Hinterbichler:2025ost}, a formally equivalent tensor appeared previously in \cite{Deser:2006zx}, where the theory of partially massless spin-2 field in de Sitter space is investigated.}
\begin{equation}\label{2.3}
f_{\mu\nu\varrho}=\partial_\mu\,h_{\nu\varrho}-\partial_\nu\,h_{\mu\varrho}.
\end{equation}

The gauge theory described by the previous action is obviously invariant under a global Poincaré symmetry, since it is written in a manifestly covariant way, and it is also invariant under the following global scaling 
\begin{equation}
x^\mu \rightarrow \sigma^{-1}\,x^\mu, \quad
\de_\mu \rightarrow \sigma\,\de_\mu, \quad
h_{\mu\nu} \rightarrow \sigma^{\frac{D-2}{2}}\,h_{\mu\nu},
\end{equation}
such that any combination of the form $\diff^D x\,\de\,h\,\de\,h$ is invariant.\footnote{In \cite{Karch:2020yuy} four models of fracton matter are considered in $2+1$ dimensions. They are obtained as the continuous limit of lattice models, involving fracton excitations due to dipole moment conservation. They are manifestly Lorentz-breaking since they involve a different number of spatial and temporal derivatives. These models are showed to enjoy subgroup of the conformal group as global symmetries.}

The equations of motion are 
\begin{align}\label{FractonEoms}
\frac{\delta S^{\text{(fr)}}}{\delta h^{\mu\nu}} &= 
-4\,\alpha\,\partial^2\,h_{\mu\nu}+(2\,\alpha-\beta)\,(\partial_\mu\,\partial_\vrho\,h^\vrho{}_\nu + \partial_\nu\,\partial_\vrho\,h^\vrho{}_\mu) \nonumber \\
& \phantom{=} + 2\,\beta\,\partial_\mu\,\partial_\nu\,h-2\,\beta\,\eta_{\mu\nu}\,(\partial^2\,h-\partial_\vrho\,\partial_\sigma\,h^{\vrho\sigma}) = 0.
\end{align} 
The case $2\,\alpha + \beta = 0$ corresponds to linearized Einstein gravity, with $h_{\mu\nu}$ viewed as the perturbation of the Minkowski metric. The gauge invariance is larger and it encompasses the full group of linearized diffeomorphisms, i.e. $\delta_\xi\,h_{\mu\nu} = \de_\mu\,\xi_\nu + \de_\nu\,\xi_\mu$. 

Let us write down the classical BV formulation of covariant fracton theories. Looking at the gauge transformation \eqref{gauge}, we can write the BRST transformation of covariant fracton theories in the sector which do not involve the antifields:
\begin{equation}\label{BRSTFracton}
s\,\lambda = 0, \quad
s\,h_{\mu\nu} = \de_\mu\,\de_\nu\,\lambda,
\end{equation}
where $s$ is the differential BRST operator, which reproduces for the gauge field $h_{\mu\nu}$ the gauge transformation with the gauge parameter $\Lambda$ promoted to an anticommuting scalar ghost $\lambda$. Denote the BV spectrum as
\begin{equation}
\{ \lambda, h_{\mu\nu}, h^*_{\mu\nu}, \lambda^* \},
\end{equation}
where $h^*_{\mu\nu}$ is the antifields of $h_{\mu\nu}$, so it is anticommuting with ghost number $-1$, and $\lambda^*$ is the antifield of $\lambda$, so it is commuting with ghost number $-2$. Summarizing, parity and ghost number of the elements in the BV spectrum are:
\begin{equation}
|\lambda|=1, \quad
|h_{\mu\nu}|=0,\quad
|h^*_{\mu\nu}|=1,\quad
|\lambda^*|=0.
\end{equation}
\begin{equation}
\mathrm{gh}(\lambda)=1,\quad
\mathrm{gh}(h_{\mu\nu})=0,\quad
\mathrm{gh}(h^*_{\mu\nu})=-1,\quad
\mathrm{gh}(\lambda^*)=-2.
\end{equation}
The BV action for covariant fracton theories is 
\begin{align}\label{BVaction}
\Gamma_{\textsc{bv}}^{\text{(fr)}} &= S^{\text{(fr)}} + \int\diff^D x\,h^{*\mu\nu}\,s\,h_{\mu\nu} = \int\diff^Dx\,(\alpha\,f^{\mu\nu\vrho}\,f_{\mu\nu\vrho} + \beta\,f_{\mu\nu}{}^\nu\,f^{\mu\vrho}{}_\vrho + h^{*\mu\nu}\,\de_\mu\,\de_\nu\,\lambda),
\end{align}
whence one can compute 
\begin{equations}
\frac{\delta_R \Gamma^{\text{(fr)}}_{\textsc{bv}}}{\delta \lambda^*} &= 0, \label{VarBV1}\\
\frac{\delta_R \Gamma^{\text{(fr)}}_{\textsc{bv}}}{\delta h^{*\mu\nu}} &=  -\de_\mu\,\de_\nu\,\lambda, \\
\frac{\delta_R \Gamma^{\text{(fr)}}_{\textsc{bv}}}{\delta h^{\mu\nu}} &= -2\,\alpha\,\de_\alpha\,f^\alpha{}_{(\mu\nu)} - 2\,\beta\,\de_\alpha\,f^{\alpha\beta}{}_\beta\,\eta_{\mu\nu} + \beta\,\de_{(\mu}\,f_{\nu)\alpha}{}^\alpha,\\
\frac{\delta_R \Gamma^{\text{(fr)}}_{\textsc{bv}}}{\delta \lambda} &=\de_\mu\,\de_\nu\,h^{*\mu\nu}.\label{VarBV4}
\end{equations}
The variation with respect to $h^{\mu\nu}$ gives the equations of motion. Using \eqref{sphisphistar} and \eqref{VarBV1}--\eqref{VarBV4}, the BV-BRST transformations of covariant fracton theories are
\begin{equations}
s\,\lambda &= 0,\label{BRSTFractonTot1}\\
s\,h_{\mu\nu} &= \de_\mu\,\de_\nu\,\lambda,\label{BRSTFractonTot2}\\
s\,h^*_{\mu\nu} &= -4\,\alpha\,\partial^2\,h_{\mu\nu}+(2\,\alpha-\beta)\,(\partial_\mu\,\partial_\lambda\,h^\lambda{}_\nu + \partial_\nu\,\partial_\lambda\,h^\lambda{}_\mu) \nonumber \\
& \phantom{=} + 2\,\beta\,\partial_\mu\,\partial_\nu\,h-2\,\beta\,\eta_{\mu\nu}\,(\partial^2\,h-\partial_\lambda\,\partial_\sigma\,h^{\lambda\sigma}), \label{BRSTFractonTot3}\\
s\,\lambda^* &= \de_\mu\,\de_\mu\,h^{*\mu\nu}.\label{BRSTFractonTot4}
\end{equations}
It will be useful in the following to notice that the integrand of the BV action \eqref{BVaction} can be written in matrix notation in the following way:
\begin{equation}\label{FractonBVMatrix}
\Gamma^{\text{(fr)}}_{\textsc{bv}} = \int\diff^D x\,
\frac{1}{2}\,\begin{pmatrix}
h^{\mu\nu} & \lambda & \lambda^* & h^{*\mu\nu}
\end{pmatrix}
\begin{pmatrix}
0 & 0 & 0 & 1 \\
0 & 0 & k & 0 \\
0 & \tilde{k} & 0 & 0 \\
2+k & 0 & 0 & 0
\end{pmatrix}
\begin{pmatrix}
s\,h_{\mu\nu} \\ s\,\lambda \\ s\,\lambda^* \\ s\,h^*_{\mu\nu}
\end{pmatrix}
\end{equation}
for any constant $k, \tilde{k}$. The ambiguity parametrized by $\tilde{k}$ is due to the fact that $s\,\lambda = 0$; The ambiguity parametrized by $k$ is due to the fact that, using \eqref{BRSTFractonTot2} and \eqref{BRSTFractonTot4}, and integrating by parts, $\lambda\,s\,\lambda^* = - h^{*\mu\nu}\,s\,h_{\mu\nu}$ modulo total derivatives.\footnote{Notice that, choosing $k = -1/\tilde{k} = -3$ (and only in this case), the matrix is symplectic.}

\restoreeqnums  

\section{Worldline models for covariant fractons}\label{worldline}

\subsection{Tensor model}\label{modelI}

The worldline tensor model is characterized by the phase space spanned by coordinates and momenta $(x^\mu,p_\mu)$ and the pair of bosonic oscillators $(\alpha^{\mu\nu},\bar{\alpha}_{\mu\nu})$, which are rank-two symmetric tensors. The symplectic structure is determined by the Poisson brackets (see Appendix~\ref{AppendixConvNot} for our conventions regarding Poisson brackets and graded commutators):
\begin{equation}\label{algebraphspI}
\{x^\mu,p_\nu\}=\delta^\mu_\nu, \quad  
\{\bar{\alpha}^{\mu\nu},\alpha_{\varrho\sigma}\}=-\tfrac{i}{2}\,(\delta^\mu_\varrho \delta^\nu _\sigma +\delta^\mu_\sigma \delta^\nu_\varrho).
\end{equation}
Consider the following set of constraints:
\begin{equation}\label{ConstraintsI}
H= p_\mu\,p^\mu, \quad 
L=\alpha^{\mu\nu} p_\mu p_\nu, \quad 
\bar{L}=\bar \alpha^{\mu\nu} p_\mu p_\nu, \quad 
J=\alpha^{\mu\nu}\bar{\alpha}_{\mu\nu}.
\end{equation} 
They form the following algebra:
\begin{equation}\label{algebraconstraintsI}
\{\bar L,L\} =-i\,H^2, \quad  
\{H,L\}=\{H,\bar L\}=0,\quad
\{J,L\}=-iL, \quad 
\{J,\bar{L}\}=i\bar L,
\end{equation}
which, remarkably, is first-class, meaning that it takes the general form \cite{VanHolten:2001nj, Newbook:2025}
\begin{equation}\label{first-class}
    \{G_a,G_b\}=P_{ab}(G),
\end{equation}
where $P_{ab}(G)$ is a homogeneous polynomial in the constraints $G_a$, with $a$ and $b$ ranging over the set of constraints, that vanishes when the constraints vanish, i.e. $P_{a b}(0)=0$. In the simplest case, the coefficients of $P_{ab}(G)$ correspond to the ``structure constants" of the constraint algebra. This is the case of \eqref{algebraconstraintsI}. However, more general scenarios are possible, as we shall encounter in the subsequent sections.

The gauged worldline action is
\begin{equation}\label{v4}
S=\int \diff\tau\,(p_\mu\,\dot x^\mu -i \bar{\alpha}^{\mu\nu}\,\dot \alpha_{\mu\nu}- e\,H - u\,\bar{L}-\bar{u}\,L - a\,J),
\end{equation}
where the ``proper time" $\tau$ is taken to range in $[0,1]$, and $e, u, \bar{u},$ and $a$ are the gauge fields for the corresponding constraints, so that the action is invariant under the gauge transformations generated by $G=\xi \,H + \varepsilon\,\bar{L} + \bar{\varepsilon}\,L +\phi\,J$ on the phase-space coordinates:
\begin{equations}
\delta x^\mu &= \{x^\mu,G\}=2\,\xi\,p^\mu + 2\,\varepsilon\bar{\alpha}^{\mu\nu}\,p_\nu+ 2\,\bar{\varepsilon}\,\alpha^{\mu\nu}\,p_\nu, \label{gaugetrI-1}\\
\delta p_\mu &=\{p_\mu,G\}=0, \\
\delta \alpha^{\mu\nu} &=\{\alpha^{\mu\nu} ,G\}=i\,\varepsilon\,p^\mu\,p^\nu + i\,\phi\,\alpha^{\mu\nu},\\
\delta \bar{\alpha}^{\mu\nu} &=\{\bar{\alpha}^{\mu\nu} ,G\}= -i\,\bar{\varepsilon}\,p^\mu\,p^\nu - i\,\phi\,\bar{\alpha}^{\mu\nu},\label{gaugetrI-2}
\end{equations}
provided that the gauge fields transform as
\begin{equations} 
\delta e &= \dot\xi+ i\,\bar{\varepsilon}\,u\,p^2 - i\,\varepsilon\,\bar{u}\,p^2,\\  
\delta u &=\dot{\varepsilon}-i\,\varepsilon\,a + i\,\phi\,u,\\ 
\delta \bar{u} &=\dot{\bar{\varepsilon}} + i\,\bar{\varepsilon}\,a-i\,\phi\,\bar{u},\\
\delta a &=\dot{\phi}.\label{gaugetrI-20}
\end{equations}
It is useful to briefly discuss the global symmetries of the worldline model. In particular, the model is invariant under the Poincar\'e group of the target space, which guarantees its relativistic invariance. If infinitesimal Lorentz transformations are parametrized by $\omega^{\mu}{}_{\nu}$, and spacetime translations by $k^\mu$, the corresponding transformations read \cite{Bastianelli:2008nm}
\begin{equation}
\delta x^\mu = \omega^\mu{}_\nu \, x^\nu + k^\mu,
\quad
\delta p_\mu = \omega_\mu{}^\nu \, p_\nu .
\end{equation}
The rank-two symmetric tensor fields $\alpha^{\mu\nu}$ and $\bar{\alpha}_{\mu\nu}$ transform as
\begin{equation}
\delta \alpha^{\mu\nu}
= \omega^\mu{}_\varrho \, \alpha^{\varrho\nu}
+ \omega^\nu{}_\varrho \, \alpha^{\mu\varrho},
\quad
\delta \bar{\alpha}_{\mu\nu}
= \omega_\mu{}^\varrho \, \bar{\alpha}_{\varrho\nu}
+ \omega_\nu{}^\varrho \, \bar{\alpha}_{\mu\varrho}.
\end{equation}
The worldline gauge fields are left invariant under both Lorentz transformations and translations. It then follows that the action~\eqref{v4} is invariant under the target-space Poincar\'e symmetry. We also note that the theory is manifestly invariant under the following global scale transformations: 
\begin{equations}
& \tau \rightarrow \sigma^p\,\tau,\quad
x^\mu\rightarrow \sigma^{-1}\,x^\mu,\quad
\alpha^{\mu\nu} \rightarrow \sigma^m\,\alpha^{\mu\nu},\\
& p_\mu \rightarrow \sigma\,p_\mu, \quad
\bar{\alpha}_{\mu\nu}\rightarrow \sigma^{-m}\,\bar{\alpha}_{\mu\nu},\\
& e\rightarrow \sigma^{-p-2}\,e,\quad
u\rightarrow \sigma^{m-p-2}\,u,\quad
\bar{u}\rightarrow \sigma^{-m-p-2}\,\bar{u},\quad
a \rightarrow \sigma^{-p}\,a,\\
& H \rightarrow \sigma^{2}\,H, \quad
\bar{L} \rightarrow \sigma^{-m-2}\,\bar{L}, \quad
L \rightarrow \sigma^{m+2}\,L, \quad
J \rightarrow J,
\end{equations}
for arbitrary numbers $m,p$.

A similar analysis applies to the worldline models discussed in the next sections and will not be repeated.

In order to quantize the system, one has to promote the canonical Poisson brackets \eqref{algebraphspI} to commutation relations:
\begin{equation}\label{commrulesI}
[x^\mu,p_\nu]=i\,\delta^\mu_\nu, \quad  
[\bar\alpha^{\mu\nu},\alpha^{\varrho\sigma}]=\tfrac{1}{2} (\eta^{\mu\varrho}\,\eta^{\nu\sigma} + \eta^{\mu\sigma}\,\eta^{\nu\varrho}).
\end{equation}
When the classical functions are promoted to operators, ordering ambiguities may arise. In particular, this is the case for the operatorial counterpart of the $J$ constraint, dubbed ``number operator". To take care of this issue, we choose the so-called ``normal ordering" for the bosonic pair ($\alpha^{\mu\nu},\bar\alpha_{\mu\nu})$, which means that the barred operators are moved to the right.

The constraints $H$, $L$, and $\bar{L}$, now treated as operators on the Hilbert space of states, satisfy the following commutation rules:
\begin{equation} \label{algebraconstraintsI@Q}
[\bar{L},L]= H^2, \quad  
[H,L]=[H,\bar{L}]=0,
\end{equation}
in correspondence with \eqref{algebraconstraintsI}. 
These constraints are implemented through the BRST charge $\mathcal{Q}$, which is defined on an enlarged Hilbert space by introducing a pair of first quantization antighost-ghost for each constraint:
\begin{align}
& H \to (b,c), \quad
L \to (B,\bar{C}), \quad
\bar{L} \to (\bar{B}, C).
\end{align}
The antighost-ghost pairs have opposite Grassmann parity with respect to the corresponding constraints. Since the constraints are all even, all the ghosts and antighosts are odd. They are taken to satisfy the following canonical commutation rule
\begin{equation}
[b,c] = 1, \quad
[B,\bar{C}] = 1, \quad
[\bar{B},C] = 1,
\end{equation}
where the graded commutator above is always an anticommutator, since the operators involved are odd. We assign the following ghost numbers for the ghosts and for the antighosts:
\begin{equation}
\mathrm{gh}(c,C,\bar{C})=+1, \quad \mathrm{gh}(b,\bar{B},B)=-1,
\end{equation}
while the remaining operators have vanishing ghost number. The commutation rules of the operators $x^\mu, p_\mu, \alpha_{\mu\nu}, \bar{\alpha}_{\mu\nu}$, $b$, $B$, $\bar{B}, c, C, \bar{C}$ can be implemented by identifying 
\begin{equation}\label{IdentificationI}
p_\mu = -i\,\de_\mu, \quad
\bar{\alpha}_{\mu\nu} = \frac{\delta}{\delta \alpha^{\mu\nu}}, \quad
b = \frac{\delta_L}{\delta c},\quad
\bar{B} = \frac{\delta_L}{\delta C}, \quad
\bar{C} = \frac{\delta_L}{\delta B},
\end{equation}
where $\frac{\delta_L}{\delta A}$ denotes the left derivative with respect to the odd variable $A$.\footnote{See also Appendix \ref{AppendixConvNot}.} To construct the BRST charge $\mathcal{Q}$, we require that it satisfies the following properties:
\begin{itemize}
\item[--] it must be an anticommuting operator $|\mathcal{Q}|=1$, with ghost number $\text{gh}(\mathcal{Q}) =+ 1$;
\item[--] it must act on the (pre-quantization) phase-space variables, upon extending the Poisson brackets to account for the ghost variables, as the gauge transformations \eqref{gaugetrI-1}--\eqref{gaugetrI-2}, with the ghost fields replacing the gauge parameters.
\item[--] it must be nilpotent: $\mathcal{Q}^2 = 0$.
\end{itemize}
These conditions enforce a basic structure that is schematically of the form: \emph{first quantization ghost $\times$ constraint}, i.e.
\begin{equation}
\mathcal{Q}= c\,H + \bar{C}\,L + C\,\bar{L} + \mathcal{M},
\end{equation}
where the first three terms directly encode the constraints, and the additional term $\mathcal{M}$ is needed to ensure nilpotency. Since the algebra \eqref{algebraconstraintsI@Q} is first-class, we can follow a well-established prescription to complete the structure and determine the last term such that the full BRST charge is automatically nilpotent \cite{VanHolten:2001nj, Fecit:2023kah, Newbook:2025}. The resulting nilpotent charge is
\begin{align}
\mathcal{Q} &= c\,H + \bar{C}\,L + C\,\bar{L} - C\,\bar{C}\,b\,H \nonumber \\
&= -c\,\de^2 - C\,\bar{\alpha}^{\mu\nu}\,\de_\mu\,\de_\nu -\alpha^{\mu\nu}\,\frac{\delta_L}{\delta B}\,\de_\mu\,\de_\nu + C\,\frac{\delta_L}{\delta B}\,\frac{\delta_L}{\delta c}\,\de^2.\label{BRSTChargeModelI}
\end{align}

In the construction above, one might wonder about the absence of a set of ghosts associated with the number operator $J$. The key idea, as first developed in \cite{Dai:2008bh} and further expanded in \cite{Bonezzi:2018box, Bonezzi:2020jjq}, is to treat it differently from the constraints triplet $(H,L,\bar L)$. Specifically, the number operator is imposed as a constraint on the BRST Hilbert space. It is within this restricted space that the cohomology of the BRST charge is to be studied. This procedure remains consistent as long as $J$ \eqref{ConstraintsI} commutes with the BRST charge. To ensure this, we extend it on the BRST Hilbert to a normal-ordered, even operator $\mathcal{J}$ with vanishing ghost number, by including the number operators associated to the ghosts, 
\begin{equation}\label{JmodelI}
\mathcal{J} = \alpha^{\mu\nu}\,\bar{\alpha}_{\mu\nu} +C \,\bar{B} + B\,\bar{C} 
= \alpha^{\mu\nu}\,\frac{\delta}{\delta \alpha^{\mu\nu}} + 
C\,\frac{\delta_L}{\delta C} + B\,\frac{\delta_L}{\delta B}, 
\end{equation}
which is indeed constructed in such a way that it commutes with the BRST charge:
\begin{equation}
[\mathcal{J},\mathcal{Q}]=0.
\end{equation}
As a consequence, $\mathcal{J}$ defines subspaces of the BRST Hilbert space, given by its eigenspaces, which are left invariant by the action of $\mathcal{Q}$. In particular, we consider the eigenspace with eigenvalue one:
\begin{equation}\label{JpsiI}
\mathcal{J}\,\psi = \psi.
\end{equation} 
The most general element $\psi$ (``string field")\footnote{We deliberately refer to an element $\psi$ of the BRST-extended Hilbert space as a ``string field". The rationale is that it plays the same role in the worldline approach as it does in string field theory: its expansion in a basis of the Hilbert space features coefficients that correspond to ordinary particle fields \cite{Zwiebach:1992ie, Gomis:1994he}.} in this subspace can be parametrized as
\begin{equation}\label{psinottracedI}
\psi =h_{\mu\nu}\,\alpha^{\mu\nu} + \lambda\,B + \tilde{\phi}\,C + \phi\,B\,c + \lambda^*\,C\,c + \tilde{h}_{\mu\nu}\,\alpha^{\mu\nu}\,c,
\end{equation}
where the operators as supposed to act on the vacuum state, which is represented by 1, according to the identification \eqref{IdentificationI}. The components $h_{\mu\nu}, \phi, \lambda, \dots$ are local tensors, which take value in the $D$-dimensional spacetime, viewed as the target space of the worldline model. In particular, $h_{\mu\nu}$ and $\tilde{h}_{\mu\nu}$ are symmetric rank-two tensors. Note that $\psi$ can be viewed as a vector 
\begin{equation}
\psi = f^i\,\psi_i,
\end{equation}
where the basis elements of the Hilbert space as listed according to the following ordering 
\begin{equation}\label{BasisModelI}
(\psi_i)_i = (\alpha^{\mu\nu}, B, C, B\,c, C\,c, \alpha^{\mu\nu}\,c),
\end{equation}
and $f^i$ denotes the field components
\begin{equation}
(f^i)_i = (h_{\mu\nu}, \lambda, \tilde{\phi},\phi,\lambda^*,\tilde{h}_{\mu\nu}).
\end{equation}
Requiring $\psi$ to be even with vanishing ghost number, the parity and the ghost number of the field components are fixed:
\begin{equation}
|h_{\mu\nu}| = |\phi| = |\lambda^*| = 0, \quad 
|\tilde{h}_{\mu\nu}| = |\lambda|= |\tilde{\phi}| = 1,
\end{equation}
\begin{equation}
\mathrm{gh}(\lambda) = 1, \quad
\mathrm{gh}(h_{\mu\nu}) = \mathrm{gh}(\phi) = 0, \quad
\mathrm{gh}(\tilde{h}_{\mu\nu}) = \mathrm{gh}(\tilde{\phi}) = -1, \quad
\mathrm{gh}(\lambda^*) = -2.
\end{equation}
This suggests that $\lambda$ could play the role of the ghost of the underlying field theory, and $\lambda^*$ on the antighost. Intuitively, one can identify $h_{\mu\nu}$ with the fracton gauge field, and $\tilde{h}_{\mu\nu}$ with the corresponding antifield. But how to identify $\phi$ and $\tilde{\phi}$ in this case? To solve the problem, we can directly operate at the level of the wordline perspective by restricting $\psi$ on a BRST-invariant smaller subspace in the Hilbert space. To do this, an even ``trace operator" with vanishing ghost number has to be introduced:
\begin{equation}
\mathcal{T}=\eta^{\mu\nu}\,\bar{\alpha}_{\mu\nu}-\bar{C}\,b = \eta^{\mu\nu}\,\frac{\delta}{\delta \alpha^{\mu\nu}} - \frac{\delta_L}{\delta B}\,\frac{\delta_L}{\delta c},
\end{equation}
which satisfies the following commutation rules 
\begin{equation}\label{commTraceI}
[\mathcal{T},\mathcal{Q}]=0, \quad [\mathcal{T},\mathcal{J}]=\mathcal{T}.
\end{equation}
Now, we can restrict $\psi$ to be an element in the kernel of $\mathcal{T}$, which in turn imposes constraints on the field components. Namely, $\phi$ should be identified with the opposite of the trace of $h_{\mu\nu}$, and $\tilde{h}_{\mu\nu}$ should be traceless:
\begin{equation}
\mathcal{T}\,\psi = 0 \Leftrightarrow 
\phi=-h_\mu{}^\mu \equiv h, \quad 
\tilde{h}_\mu{}^\mu = 0.
\end{equation}
Replacing in \eqref{psinottracedI}, we obtain 
\begin{equation}\label{stringfieldI}
\hat{\psi} = h_{\mu\nu}\,\alpha^{\mu\nu} +\lambda\,B + \tilde{\phi}\,C - h\,B\,c + \lambda^*\,C\,c + \hat{\tilde{h}}_{\mu\nu}\,\alpha^{\mu\nu}\,c = \hat{f}^i\,\hat\psi_i, \quad\text{such that}\;\mathcal{T}\,\hat{\psi} = 0,
\end{equation}
where $\hat{\tilde{h}}_{\mu\nu}$ is the traceless part of $\tilde{h}_{\mu\nu}$, and, according to the ordering \eqref{BasisModelI}, the field components are
\begin{equation}
(\hat{f}^i)_i = (h_{\mu\nu},\lambda,\tilde{\phi},-h,\lambda^*,\hat{\tilde{h}}_{\mu\nu}).
\end{equation} 
The BRST charge acts on $\hat\psi$ according to 
\begin{align}
\mathcal{Q}\,\hat{\psi} &=\de_\mu\,\de_\nu\,\lambda\,\alpha^{\mu\nu} + (\de^2\,h-\de_\mu\,\de_\nu\,h^{\mu\nu})\,C - \de^2\,\lambda\,B\,c \nonumber\\
&\phantom{=}+ (\de_\mu\,\de_\nu\,\hat{\tilde{h}}^{\mu\nu}-\de^2\,\tilde{\phi})\,C\,c + (\de_\mu\,\de_\nu\,h - \de^2\,h_{\mu\nu})\,\alpha^{\mu\nu}\,c.\label{QpsiModelI}
\end{align}

Once the BRST charge $\mathcal{Q}$ is introduced and a subsector of the BRST Hilbert space, closed under the $\mathcal{Q}$-action, is identified, one typically proceeds by studying $\mathcal{Q}$-cohomology within this subsector. In our case, this amounts to solving the equation $\mathcal{Q}\,\hat\psi = 0$ modulo $\mathcal{Q}$-exact states $\hat\psi \sim \hat\psi + \mathcal{Q}\,\Lambda$. At ghost number zero, the field components of the state $\hat\psi$ in the $\mathcal{Q}$-kernel satisfy the field equations, which correspond to the equations of motion of the underlying spacetime theory. The $\mathcal{Q}$-exactness amounts to the gauge redundancy.

The worldline formalism, however, provides a more powerful framework, as it allows one to derive the BRST transformations of the whole BV spectrum, encompassed in $\hat\psi$. In particular, the transformation of the antifield, which occupies the ghost number zero sector, corresponds to the equations of motion. The precise relation between the first-quantized $\mathcal{Q}$, acting on $\hat\psi$, and the BRST differential operator $s$, acting on the field components of $\hat\psi$, is \cite{Bonezzi:2024emt}
\begin{equation}\label{BRSTdifferential}
\mathcal{Q}\,\hat{\psi} \;=\; s\,\hat{f}^i\,\hat{\psi}_i \,.
\end{equation}
Consistently, $s$ is odd, nilpotent with ghost number one by construction:
\begin{equation}
s^2 = 0, \quad \mathrm{gh}(s) = 1, \quad |s|=1.
\end{equation}
Using \eqref{QpsiModelI}, one finds 
\begin{equations}
s\,\lambda &= 0\label{BRSTFractonModelI1},\\
s\,h_{\mu\nu} &= \partial_\mu\,\partial_\nu\,\lambda,\label{BRSTFractonModelI2}\\
s\,h &= \partial^2\,\lambda,\label{BRSTFractonModelI2bis}\\
s\,\hat{\tilde{h}}_{\mu\nu} &= \partial_\mu\,\partial_\nu\,h - \partial^2\,h_{\mu\nu},\label{BRSTFractonModelI3}\\
s\,\tilde{\phi} &= \partial^2\,h-\partial_\mu\partial_\nu\,h^{\mu\nu},\label{BRSTFractonModelI4}\\
s\,\lambda^* &= \partial_\mu\,\partial_\nu\,\hat{\tilde{h}}^{\mu\nu}-\partial^2\,\tilde{\phi}.\label{BRSTFractonModelI5}
\end{equations}
Notice that the transformation \eqref{BRSTFractonModelI3} is traceless, as it should be. The transformation \eqref{BRSTFractonModelI2bis} is implied by \eqref{BRSTFractonModelI2}, so that it can be discharged. The  transformations \eqref{BRSTFractonModelI1} and \eqref{BRSTFractonModelI2} are the same as in covariant fracton theories \eqref{BRSTFracton} or \eqref{BRSTFractonTot1}-\eqref{BRSTFractonTot2}. Therefore, $h_{\mu\nu}$ can be identified with the fracton gauge field, and $\lambda$ with the fracton ghost, as the notation suggests. The remaining transformations are equivalent to the BRST transformations of the antifield sector of the theory \eqref{BRSTFractonTot3}--\eqref{BRSTFractonTot4} in the peculiar case 
\begin{equation}\label{ConstModelI}
2\alpha -\beta=0,
\end{equation}
with the normalization $\alpha = \tfrac{1}{4}$, so that $\beta = \tfrac{1}{2}$. Indeed, using the following change of variables 
\begin{equation}\label{ChangeOfVariablesModelI}
\hat{\tilde{h}}_{\mu\nu} = h^*_{\mu\nu} -\tfrac{1}{D}\,\eta_{\mu\nu}\,h^*_\vrho{}^\vrho, \quad
\tilde{\phi} = -\tfrac{1}{D}\,h^*_\vrho{}^\vrho,
\end{equation}
the transformations \eqref{BRSTFractonModelI1}--\eqref{BRSTFractonModelI5} become
\begin{equations}
s\,\lambda &= 0,\\
s\,h_{\mu\nu} &= \de_\mu\,\de_\nu\,\lambda,\\
s\,h^*_{\mu\nu} &= -\de^2\,h_{\mu\nu} + \de_\mu\,\de_\nu\,h - \eta_{\mu\nu}\,(\de^2\,h - \de_\vrho\,\de_\sigma\,h^{\vrho\sigma}),\\
s\,\lambda^* &= \de_\mu\,\de_\nu\,h^{*\mu\nu},
\end{equations}
which are consistent with \eqref{BRSTFractonTot1}--\eqref{BRSTFractonTot4}, upon setting \eqref{ConstModelI}.

Let us comment on the fact that one might have expected that the model under examination would reproduce the specific case of fractonic gauge theory, identified by \eqref{ConstModelI}, right from the beginning. Indeed, the constraint structure in \eqref{ConstraintsI} does not allow for the vector divergence $\de^\mu\,h_{\mu\nu}$ to be reproduced. This contraction precisely drops out from the action \eqref{action} and from the equations of motion \eqref{FractonEoms} equations only in the case \eqref{ConstModelI}.

To go beyond this limitation, it seems necessary to consider constraints in which the two momenta -- both of which are required in the construction, as discussed in the introduction -- are not simultaneously contracted with the same bosonic oscillator. We will leave this analysis for the next section.

Akin to string field theory \cite{Siegel:1988yz, Zwiebach:1992ie}, it is possible to obtain the action of the field theory using the worldline formulation. The action corresponds to the spacetime integral of the inner product in the Hilbert space between the string field $\psi$ and its BRST variations $\mathcal{Q}\,\psi$. Although the inner product in the Hilbert space can be defined algebraically in a completely abstract way \cite{Barnich:2003wj, Bengtsson:2004cd, Barnich:2004cr, Grigoriev:2006tt}, it is somehow useful to have in mind an explicit representation. We can use the following one:\footnote{See also Appendix \ref{AppendixConvNot}.}
\begin{equation}\label{InnerProductI}
\braket{\psi,\vphi} = i\int\diff c\,\bar{\psi}\,\vphi,
\end{equation}
for any element $\psi,\vphi$ in the Hilbert space. The integral is the Berezin integral over the odd variable $c$, which is assumed to be real. This inner product is odd with ghost number $-1$, as a consequence of $\int\diff c\,c = 1$. It implies the following normalization
\begin{equation}
\braket{1,c} = i,
\end{equation}
and it satisfies the following property
\begin{equation}
\overline{\braket{\psi,\vphi}} = (-)^{|\psi|+|\vphi|}\,\braket{\vphi,\psi}, 
\end{equation}
where the overline denotes the complex conjugation. The inner product allows us to define the adjoint of an operator:
\begin{equation}
\braket{\psi,A^\dagger\,\vphi} = \braket{A\,\psi,\vphi}.
\end{equation}
The BRST charge defined in \eqref{BRSTChargeModelI} is self-adjoint with respect to the inner product
\begin{equation}
\braket{\psi,\mathcal{Q}\,\vphi} = \braket{\mathcal{Q}\,\psi,\vphi}, \quad \forall\,\psi,\vphi,
\end{equation}
provided that 
\begin{equation}\label{HermitianCond1}
\alpha_{\mu\nu}^\dagger = \bar{\alpha}_{\mu\nu}, \quad
c^\dagger\ = c, \quad 
C^\dagger = \bar{C} = \frac{\delta_L}{\delta B},
\end{equation}
which, together with the commutation rules \eqref{BasisModelI}, imply that
\begin{equation}\label{HermitianCond2}
b^\dagger = b = \frac{\delta_L}{\delta c}, \quad B^\dagger = \bar{B} = \frac{\delta_L}{\delta C}.
\end{equation}
The unique non-vanishing products between two elements in the basis \eqref{BasisModelI} are
\begin{align}
& \braket{\alpha_{\mu\nu},\alpha^{\vrho\sigma}\,c} =
\braket{\alpha^{\vrho\sigma}\,c,\alpha_{\mu\nu}} = 
\tfrac{i}{2}\,\delta^\vrho_{(\mu}\,\delta^\sigma_{\nu)}, \\
& \braket{B,C\,c} = \braket{C\,c,B} = \braket{C,B\,c} = \braket{B\,c,C} = i,
\end{align}
so that, using the ordering in \eqref{BasisModelI}, we can write the following matrix:
\begin{equation}
(\braket{\psi_i,\psi_j})_{ij} = \begin{pmatrix}
0 & 0 & 0 & 0 & 0 & \frac{i}{2}\,\delta_{(\mu}^\vrho\,\delta_{\nu)}^\sigma \\
0 & 0 & 0 & 0 & i & 0 \\
0 & 0 & 0 & i & 0 & 0 \\
0 & 0 & i & 0 & 0 & 0 \\
0 & i & 0 & 0 & 0 & 0 \\
\frac{i}{2}\,\delta_{(\mu}^\vrho\,\delta_{\nu)}^\sigma & 0 & 0 & 0 & 0 & 0
\end{pmatrix}.
\end{equation}

Observe that
\begin{equation}
\braket{\psi,Q\,\psi} = \braket{f^i\,\psi_i,s\,f^j\,\psi_j} = \braket{\psi_i,\psi_j}\,f^i\,s\,f^j = f^i\,\braket{\psi_i,\psi_j}\,s\,f^j.
\end{equation}
Indeed, $f^i$ is real, and both $\braket{\psi_i,\psi_j}$ and $f^i\,s\,f^j$ are even.\footnote{If $\psi_i$ is even/odd, $f^i$ is even/odd, since $\psi$ is even; if $\psi_i$ is even/odd, then $\psi_j$ is odd/even, in order the inner product not to vanish; if $\psi_j$ is odd/even, then $s\,f^j$ is even/odd, since $Q\,\psi$ is odd.} Therefore, \emph{if we suppose $\lambda^*$ to be pure imaginary and the other fields to be real},\footnote{Actually, $\lambda$ could be real or imaginary, because $s\,\lambda = 0$.} we can write in matrix notation
\begin{equation}
-i\,\braket{\hat{\psi},Q\,\hat{\psi}} = 
\begin{pmatrix}
h^{\mu\nu} & -\lambda & \tilde{\phi} & -h & \lambda^* & \hat{\tilde{h}}^{\mu\nu}
\end{pmatrix}
\begin{pmatrix}
0 & 0 & 0 & 0 & 0 & 1 \\
0 & 0 & 0 & 0 & 1 & 0 \\
0 & 0 & 0 & 1 & 0 & 0 \\
0 & 0 & 1 & 0 & 0 & 0 \\
0 & 1 & 0 & 0 & 0 & 0 \\
1 & 0 & 0 & 0 & 0 & 0 
\end{pmatrix}
\begin{pmatrix}
s\,h_{\mu\nu} \\ s\,\lambda \\ s\,\tilde{\phi} \\ -s\,h \\ s\,\lambda^* \\ s\,\hat{\tilde{h}}_{\mu\nu}
\end{pmatrix},
\end{equation}
Equivalently, using $h^*_{\mu\nu}$ in place of $\hat{\tilde{h}}_{\mu\nu}$ and $\tilde{\phi}$, according to \eqref{ChangeOfVariablesModelI}, it becomes
\begin{equation}
-i\,\braket{\hat{\psi},Q\,\hat{\psi}} =  \begin{pmatrix}
h^{\mu\nu} & \lambda & \lambda^* & h^{*\mu\nu}
\end{pmatrix}
\begin{pmatrix}
0 & 0 & 0 & 1 \\
0 & 0 & -1 & 0 \\
0 & -1 & 0 & 0 \\
1 & 0 & 0 & 0
\end{pmatrix}
\begin{pmatrix}
s\,h_{\mu\nu} \\ s\,\lambda \\ s\,\lambda^* \\ s\,h^*_{\mu\nu}
\end{pmatrix},
\end{equation}
which corresponds to \eqref{FractonBVMatrix}, choosing $k=\tilde{k}=-1$. Therefore, we conclude that
\begin{equation}
\Gamma^{\text{(fr)}}_{\textsc{bv}}(\alpha=\tfrac{1}{4},\beta=\tfrac{1}{2}) = -\frac{i}{2}\int\diff^D x\,\braket{\hat\psi,\mathcal{Q}\,\hat\psi}.
\end{equation}

\subsection{Vector model}\label{modelII}

In this section, we study another worldline model, which again it is shown to reproduce a covariant fracton gauge theory. At the classical level, the phase space is spanned by coordinates and momenta $(x^\mu,p_\mu)$ and by a pair of bosonic oscillators $(\alpha^{\mu},\bar\alpha_{\mu})$, which, unlike the model studied in the previous section, carry a single spacetime index. The symplectic structure is 
\begin{equation}\label{algebraphspII}
\{x^\mu,p_\nu\}=\delta^\mu_\nu\;, \quad  \{\bar{\alpha}^{\mu},\alpha_{\nu}\}=-\,i\,\delta^\mu_\nu.
\end{equation}
Consider the following constraints:
\begin{equation}\label{constraintsII}
\ell = \alpha^\mu\,p_\mu\,\bar{\alpha}^\nu\,p_\nu, \quad 
L=(\alpha^{\mu} p_\mu)^2, \quad 
\bar{L}= (\bar{\alpha}^{\mu}\,p_\mu)^2, \quad 
J= \tfrac{1}{2}\,\alpha^{\mu}\,\bar{\alpha}_{\mu}.
\end{equation} 
They satisfy
\begin{equation}\label{algebraconstraintsII}
\{\bar{L},L\} = - 4\,i\,H\,\ell,\quad  
\{\ell,L\} = -2\,i\,H\,L, \quad 
\{\ell,\bar L\} = 2\,i\,H\,\bar{L},
\end{equation}
\begin{equation}\label{algebraconstraintsII.2}
\{J,L\}= -i\,L, \quad \{J,\bar L\}= i\,\bar{L}, \quad \{J,\ell\} =0,
\end{equation}
where $H=p_\mu\,p^\mu $. As in the model discussed in the previous section, this is a first-class algebra, as it takes the general form \eqref{first-class}. However, in this case, the polynomial $P_{ab}(G)$ features ``structure functions" as coefficients, that is, coefficients that depend on the phase-space variables. This dependence is evident in the appearance of $H$ in the first line above. Unlike in the previous model, $H$ is \emph{not} treated as a constraint here,\footnote{One may wonder what happens when the phase-space function $H$ is also treated as a constraint. Proceeding with BRST quantization and including $H$ as an independent constraint requires reintroducing the pair of first-quantization ghosts $(b,c)$ associated with the Hamiltonian, so that it can be included in the BRST charge. This modification leads to two copies of the same spectrum described in this section: the BRST Hilbert space is effectively doubled by the presence of the $c$ ghost. Consequently, the resulting equations split into two equivalent sets, differing only by a shift in ghost number and a change in the Grassmann parity of the fields, but otherwise physically equivalent.} and the coefficients are therefore not constant numerical values. The gauged worldline action is
\begin{equation}
S=\int \diff\tau\,(p_\mu\,\dot x^\mu -i\,\bar\alpha^\mu\, \dot\alpha_{\mu}- \omega\,\ell- u\,\bar{L} - \bar{u}\,L - a\,J),
\end{equation}
having introduced the set of Lagrange multipliers $(\omega,u,\bar u,a)$. Given the generator of gauge transformations $G=\chi \,\ell + \varepsilon\,\bar{L} + \bar{\varepsilon}\,L + \phi\,J$, the phase-space variables transform according to
\begin{equations}
\delta x^\mu &= \{x^\mu,G\}=\chi\,\alpha^\mu \,p_\nu\,\bar\alpha^\nu + \chi\,\bar\alpha^\mu\,p_\nu\,\alpha^\nu + 2\,\varepsilon\,\bar{\alpha}^\mu\,p_\nu\,\bar\alpha^\nu+ 2\,\bar\varepsilon\,\alpha^\mu\,p_\nu\,\alpha^\nu,\label{gaugetrsII1}\\
\delta p_\mu &=\{p_\mu,G\}=0, \\
\delta \alpha^{\mu} &=\{\alpha^{\mu} ,G\} = i\,\chi\,p^\mu\,p_\nu\,\alpha^\nu + 2\,i\,\varepsilon\,p^\mu\,p_\nu\,\bar\alpha^\nu + \tfrac{i}{2}\,\phi\,\alpha^{\mu},\\
\delta \bar{\alpha}^\mu &=\{\bar\alpha^{\mu},G\}=-i\,\chi\,p^\mu\,p_\nu\,\bar\alpha^\nu-2\,i\,\bar\varepsilon\,p^\mu\,p_\nu\,\alpha^\nu-\tfrac{i}{2}\,\phi\,\bar\alpha^\mu,
\end{equations}
and the action is invariant, provided that
\begin{equations} 
\delta \omega &= \dot\chi + 4\,i\,H\,\bar\varepsilon\,u - 4\,i\,H\,\varepsilon\,\bar u,\\  
\delta u &= \dot{\varepsilon} + 2\,i\,H\,\chi\,u - 2\,i\,H\,\varepsilon \,\omega - i\,\varepsilon\,a + i\,\phi\,u,\\ 
\delta \bar u &= \dot{\bar\varepsilon}- 2\,i\,H\,\chi \,\bar{u} + 2\,i\,H\,\bar\varepsilon\,\omega+ i\,\bar\,\varepsilon a - i\,\phi\,\bar u,\\
\delta a &=\dot{\phi}.\label{gaugetrsII2}
\end{equations}
Upon quantization, we promote the canonical Poisson brackets \eqref{algebraphspII} to commutation relations
\begin{equation}
[x^\mu,p_\nu]=i\,\delta^\mu_\nu, \quad  
[\bar{\alpha}^{\mu},\alpha^{\nu}]=\eta^{\mu\nu}.
\end{equation}
We then proceed with the BRST quantization. Once again, we have to deal with ordering ambiguities: in this case, we adopt the \emph{symmetric} or \emph{Weyl ordering} for the bosonic-oscillators sector.\footnote{Given two classical functions $a$ and $b$, when they are promoted to non-commuting operators $\hat{a}$ and $\hat{b}$, the symmetric-ordered quantization of the product $a\,b$ is  $\frac{1}{2}\,(\hat{a}\,\hat{b} + \hat{a}\,\hat{b})$, which is not equivalent to $\hat{a}\,\hat{b}$, since $\frac{1}{2}\,(\hat{a}\,\hat{b} + \hat{a}\,\hat{b}) = \hat{a}\,\hat{b} - \tfrac{1}{2}\,[\hat{a},\hat{b}]$.} In particular, we are led to define the following quantum operator:
\begin{equation}\label{SymmetricEll}
\ell_{\mathrm{w}} = \alpha^\mu\,p_\mu\,\bar{\alpha}^\nu\,p_\nu + \tfrac{1}{2}\,H,
\end{equation}
where the appearance of the second term is a quantum ordering effect. As a consequence of this ordering prescription, the algebra formally maintains the same structure as its classical counterpart:
\begin{equation}\label{algebraconstraintsII@Q}
[\bar L,L] = 4\,H\,\ell_{\mathrm{w}}, \quad  
[\ell_{\mathrm{w}},L]=2\,H\,L, \quad 
[\ell_{\mathrm{w}},\bar L]=-2\,H\,\bar L,
\end{equation}
\begin{equation}\label{algebraconstraintsII@Q.2}
[J,L]=L, \quad [J,\bar L]=\bar L, \quad [J,\ell_{\mathrm{w}}] =0.
\end{equation}

In defining the BRST Hilbert space, the pair of first-quantized antighost-ghost for the constraints $L$ and $\bar{L}$ is the same as in the previous section. Since the remaining constraint is now $\ell$ instead of $H$, we consider a new pair $(\mathfrak{b},\mathfrak{c})$, but parity, ghost number, commutation rule, and identification with operator derivative are the same as those of the pair $(b,c)$, associated to $H$ in the previous section:
\begin{equation}\label{HilbertModelII}
\ell_{\mathrm{w}} \rightarrow (\mathfrak{b},\mathfrak{c}), \quad
[\mathfrak{b},\mathfrak{c}] = 1, \quad |\mathfrak{b}|=|\mathfrak{c}| = 1, \quad \text{gh}(\mathfrak{b}) = -1, \quad \text{gh}(\mathfrak{c}) =+ 1, \quad \mathfrak{b} = \frac{\delta_L}{\delta\mathfrak{c}}.
\end{equation}
In a sense, the constraint $\ell$ plays a role in this model analogous to that of the Hamiltonian $H$ in model discussed in the previous section; this can also be inferred from the similarities between the gauge transformations \eqref{gaugetrsII1}--\eqref{gaugetrsII2} generated by $\ell$ and those of the previous model \eqref{gaugetrI-1}--\eqref{gaugetrI-20} generated by $H$ on the phase-space variables and on the gauge fields.\footnote{One may wonder whether not-gauging the Hamiltonian constraint in a worldline model could lead to issues, as it certainly would in ordinary particle models. In the present case, we do not have a definite answer, and a deeper analysis, most likely involving the construction of the path integral, would be needed. Nonetheless, it would not be surprising to encounter instabilities already at the classical level, as this is a recurring issue in covariant fracton theories, even from a field-theoretic perspective \cite{Rovere:2024nwc, Afxonidis:2024tph}.} Choosing once again the normal ordering for the ghost sector, the nilpotent BRST charge is
\begin{align}\label{QII}
\mathcal{Q} &= \mathfrak{c}\,\ell_{\mathrm{w}} + \bar{C}\,L + C\,\bar{L} - 4\,C\, \bar C\,\mathfrak{b}\,H - 2\,\bar{C}\,B\,\mathfrak{c}\,H + 2\,C\,\bar{B}\,\mathfrak{c}\,H  \nonumber \\
&= \tfrac{3}{2}\,\mathfrak{c}\,\de^2 - \mathfrak{c}\,\alpha^\mu\,\tfrac{\delta_L}{\delta\alpha_\nu}\,\de_\mu\,\de_\nu -\alpha^\mu\,\alpha^\nu\,\tfrac{\delta_L}{\delta B}\,\de_\mu\,\de_\nu - C\,\tfrac{\delta_L}{\delta\alpha_\mu}\,\tfrac{\delta}{\delta\alpha_\nu}\,\de_\mu\,\de_\nu \nonumber \\
& \phantom{=} + 4\,C\,\tfrac{\delta_L}{\delta B}\,\tfrac{\delta_L}{\delta\mathfrak{c}}\,\de^2 - 2\,\mathfrak{c}\,B\,\tfrac{\delta_L}{\delta B}\,\de^2 - 2\,\mathfrak{c}\,C\,\tfrac{\delta_L}{\delta C}\,\de^2.
\end{align}

Let us pause for a moment to comment on the consequences of the choice of operator ordering. As previously emphasized, starting from the classical expression for the constraint $\ell$, and adopting symmetric ordering for the bosonic oscillators, leads to the expression \eqref{SymmetricEll}, which ensures that the constraint algebra remains first-class even at the quantum level, as in \eqref{algebraconstraintsII@Q}--\eqref{algebraconstraintsII@Q.2}. The standard BRST construction can be directly applied to obtain a nilpotent charge. 
Consider now a different ordering prescription, for example, the normal ordering. If one starts from the same classical constraint $\ell$ and quantizes it using normal ordering, then the quantum operator is 
\begin{equation}\label{lN}
\ell_{\mathrm{N}} = \alpha^\mu p_\mu \bar\alpha^\nu p_\nu,
\end{equation}
and one can check that the resulting algebra is no longer first-class. Nevertheless, one can still construct a nilpotent BRST charge by adding a suitable $\sim\mathfrak{c} \,H$ term by hand, getting the same expression for the nilpotent charge as in the symmetric ordering case \eqref{QII}. However, such an ad hoc term lacks a natural justification at the level of constraints, since $H$ is not classically a constraint.

We can then define an operator $\mathcal{J}$, on the same footing as \eqref{JmodelI}, 
\begin{equation}
\mathcal{J}= \frac{1}{2}\,\alpha^{\mu}\,\bar\alpha_{\mu} + C\,\bar{B} + B\,\bar{C} +\frac{D}{4} = \frac{1}{2}\,\alpha^\mu\,\frac{\delta}{\delta\alpha^\mu} + C\,\frac{\delta_L}{\delta C} + B\,\frac{\delta_L}{\delta B} +\frac{D}{4},
\end{equation}
where the $\tfrac{D}{4}$ shift is a consequence of the Weyl-ordering prescription. This operator commutes with $\mathcal{Q}$, thus we are free to consider the most general element in the eigenspace of $\mathcal{J}$ with eigenvalue $\tfrac{D+4}{4}$:
\begin{equation}\label{psinottracedII}
\psi = h_{\mu\nu}\,\alpha^\mu\,\alpha^\nu + \lambda\,B + \tilde{\phi}\,C + \phi\,B\,\mathfrak{c} + \tilde{\lambda}\,C\,\mathfrak{c} + \tilde{h}_{\mu\nu}\,\alpha^\mu\,\alpha^\nu\,\mathfrak{c} = f^i\,\psi_i,
\end{equation}
where the basis elements are
\begin{equation}
(\psi_i)_i = (\alpha^\mu\,\alpha^\nu,B,C,B\,\mathfrak{c},C\,\mathfrak{c},\alpha^\mu\,\alpha^\nu\,\mathfrak{c}),
\end{equation}
and the field components are 
\begin{equation}
(f^i)_i = (h_{\mu\nu},\lambda,\tilde{\phi},\phi,\tilde{\lambda},\tilde{h}_{\mu\nu}).
\end{equation}
The parity and ghost numbers of the latter are fixed by requiring $\psi$ to be even with vanishing ghost number:
\begin{equation}
|h_{\mu\nu}| = |\phi| = |\tilde{\lambda}|=0, \quad
|\lambda|=|\tilde{\phi}|=|\tilde{h}_{\mu\nu}|= 1,
\end{equation}
\begin{equation}
\text{gh}(\lambda) = 1, \quad
\text{gh}(h_{\mu\nu}) = \text{gh}(\phi) = 0, \quad
\text{gh}(\tilde{h}_{\mu\nu}) = \text{gh}(\tilde{\phi}) = -1, \quad
\text{gh}(\tilde{\lambda}) = -2.
\end{equation}
As in the model of Section~\ref{modelI}, we introduce a ``trace operator", in order to constrain $\phi$ and $\tilde{\phi}$, which are redundant in comparison with the BV spectrum of covariant fracton gauge theories:
\begin{equation}\label{TraceII}
\mathcal{T}= \tfrac{1}{2}\,\bar\alpha^\mu\,\bar\alpha_\mu+ 2\,\mathfrak{b}\,\bar{C}-\mathfrak{c}\,\bar{B} = \frac{1}{2}\,\frac{\delta}{\delta\alpha^\mu}\,\frac{\delta}{\delta\alpha_\mu} + 2\,\frac{\delta_L}{\delta\mathfrak{c}}\,\frac{\delta_L}{\delta B} + \mathfrak{c}\,\frac{\delta_L}{\delta C}.
\end{equation}
One can check that $\mathcal{T}$ enjoys the same commutation rules, cf. \eqref{commTraceI}, as the corresponding operator in Section~\ref{modelI}. Requiring $\psi$ to sit in the kernel of $\mathcal{T}$, one gets the following two conditions:
\begin{equation}
\phi= -\frac{1}{2}\,h_\mu{}^\mu = -\frac{1}{2}\,h, \quad \tilde{\phi} = \tilde{h}_\mu{}^\mu = \tilde{h},
\end{equation}
so that we consider the state
\begin{equation}\label{psiII}
\hat{\psi} =  h_{\mu\nu}\,\alpha^\mu\,\alpha^\nu + \lambda\,B + \tilde{h}\,C -\frac{1}{2}\,h\,B\,\mathfrak{c} + \tilde{\lambda}\,C\,\mathfrak{c} + \tilde{h}_{\mu\nu}\,\alpha^\mu\,\alpha^\nu\,\mathfrak{c} = \hat{f}^i\,\hat\psi_i.
\end{equation}
The BRST charge acts on $\hat\psi$ according to 
\begin{align}
\mathcal{Q}\,\hat{\psi} &=\de_\mu\,\de_\nu\,\lambda\,\alpha^{\mu}\,\alpha^{\nu} +2\,(\de^2\,h-\de_\mu\,\de_\nu\,h^{\mu\nu})\,C - \tfrac12\,\de^2\,\lambda\,B\,\mathfrak{c} \nonumber\\
&\phantom{=} + 2\,(\de_\mu\,\de_\nu\,\tilde{h}^{\mu\nu}-\tfrac14\,\de^2\,\tilde{h})\,C\,\mathfrak{c} + (\tfrac{3}{2}\,\partial^2\,h_{\mu\nu}-\partial^\vrho\,\partial_{(\mu}\,h_{\nu)\vrho} +\tfrac{1}{2}\partial_\mu\,\partial_\nu\,h)\,\alpha^{\mu}\,\alpha^{\nu}\,\mathfrak{c},\label{QpsiModelII}
\end{align}
and using $\mathcal{Q}\,\hat{\psi} = s\,\hat{f}^i\,\hat\psi_i$, one can compute the BRST transformations on the field components:
\begin{equations}
s\,\lambda & =0,\label{eomsII1}\\
s\,h_{\mu\nu} &=\partial_\mu\,\partial_\nu\,\lambda,\label{eomsII1a}\\
s\,h &= \partial^2\,\lambda,\label{eomsII1b}\\
s\,\tilde{h}_{\mu\nu} &=\tfrac{3}{2}\,\partial^2\,h_{\mu\nu}-\partial^\vrho\,\partial_{(\mu}\,h_{\nu)\vrho} +\tfrac{1}{2}\,\partial_\mu\,\partial_\nu\,h, \label{eomsII1c}\\
s\,\tilde{h} &= 2\,\partial^2\,h-2\,\partial_\mu\,\partial_\nu\,h^{\mu\nu},\label{eomsII1d}\\
s\,\tilde{\lambda} &=-\tfrac12\,\partial^2\,\tilde{h} + 2\,\partial_\mu \,\partial_\nu\,\tilde{h}^{\mu\nu}.\label{eomsII2}
\end{equations}
The transformations \eqref{eomsII1} and \eqref{eomsII1a} are respectively the expected BRST transformations for ghost and gauge field of covariant fracton theories; the transformations \eqref{eomsII1b} and \eqref{eomsII1d} are the trace of \eqref{eomsII1a} and of \eqref{eomsII1c} respectively, so that they can be discharged. Now, using the following change of variables 
\begin{equation}
\tilde{h}_{\mu\nu} = h^*_{\mu\nu} + \tfrac{1}{D-4}\,\eta_{\mu\nu}\,h^*, \quad 
\tilde{\lambda} = 2\,\lambda^*,
\end{equation}
which is valid when $D \neq 4$ (we will comment on this peculiar case later), and replacing in the independent BRST transformations \eqref{eomsII1}--\eqref{eomsII2}, we obtain
\begin{equations}
s\,\lambda &= 0,\\
s\,h_{\mu\nu} &= \de_\mu\,\de_\nu\,\lambda,\\
s\,h^*_{\mu\nu} &= \tfrac{3}{2}\,\de^2\,h_{\mu\nu} + \tfrac{1}{2}\,\de_\mu\,\de_\nu\,h - \de^\vrho\,\de_{(\mu}\,h_{\nu)\vrho} + \tfrac{1}{2}\,\eta_{\mu\nu}\,(\de^2\,h - \de_\vrho\,\de_\sigma\,h^{\vrho\sigma}),\\
s\,\lambda^* &= \de_\mu\,\de_\nu\,h^{*\mu\nu}.
\end{equations}
Comparing with \eqref{BRSTFractonTot1}--\eqref{BRSTFractonTot4}, we find that they correspond to the BRST rules in the case
\begin{equation}\label{ConstModelII}
2\,\alpha + 3\,\beta = 0,
\end{equation}
with the normalization choice 
\begin{equation}\label{ConstModelIIbis}
\alpha = -\tfrac{3}{8}, \quad \beta = \tfrac{1}{4}.
\end{equation}
We observe that the equations of motion of the theory \eqref{FractonEoms} are traceless in $D = 4$, with this choice of the constants. Indeed, the trace equation reads
\begin{equation}
\eta^{\mu\nu}\,\frac{\delta S^{\text{(fr)}}}{\delta h^{\mu\nu}} = 2\,(2\,\alpha + (d-1)\,\beta)\,(\de_\mu\,\de_\nu\,h^{\mu\nu} - \de^2\,h),
\end{equation}
so that\footnote{In the traceless theory, the BRST symmetry is enlarged, because the theory is also invariant under a conformal ``scaling" $ \delta_\sigma\,h_{\mu\nu} = 2\,\sigma\,\eta_{\mu\nu}$. Indeed, the transformation of the fracton field strength is $\delta_\sigma\,f_{\mu\nu\vrho} = 2\,\de_{[\mu}\,\sigma\,\eta_{\nu]\vrho}$, so that one can compute the scaling of the Lagrangian to be equal to $\delta_\sigma\,(\alpha\,f_{\mu\nu\vrho}\,f^{\mu\nu\vrho} + \beta\,f^{\mu\nu}{}_\nu\,f_{\mu\vrho}{}^\vrho) = 4\,(2\,\alpha + (D-1)\,\beta)\,f^{\mu\nu}{}_\nu\,\de_\mu\,\sigma$, which allows to conclude that the theory is scaling-invariant if and only if the condition \eqref{TracelessCond} is met.}
\begin{equation}\label{TracelessCond}
\text{The theory is traceless} \Leftrightarrow  2\,\alpha + (D-1)\,\beta = 0,
\end{equation}
and, by plugging  \eqref{ConstModelIIbis} in the previous condition, one finds $D=4$. This case is not captured by the worldline model discussed in this section, since it produces a trace equation also in the case $D=4$.

Let us comment on the reason why the wordline vector model discussed in this section describes the peculiar covariant fracton theory, described by the choice of the constants such that \eqref{ConstModelII} is fulfilled. We have already emphasized the special role played by the combination of $\ell_\mathrm{N} = \alpha^\mu p_\mu \bar\alpha^\nu p_\nu$ and $H$ in \eqref{SymmetricEll}, especially in relation to the quantum algebra and the construction of a nilpotent BRST operator. As previously discussed, this specific combination arises naturally when employing the Weyl-ordered quantum constraint $\ell_{\mathrm{w}}$. The selection of the fracton theory \eqref{ConstModelII} is a consequence of this distinguished combination, since the precise form of the quantum operator $\ell_{\mathrm{w}}$ inevitably fixes a specific ratio between the coefficients of the Laplacian term $\partial^2 h_{\mu\nu}$ and the divergence term $\partial_\varrho\,\partial_\mu\,h_\nu{}^\varrho$ in the BRST transformations. This ratio is uniquely realized when the parameters $\alpha,\beta$ satisfy the constraint \eqref{ConstModelII}. Thus, it is not the field theory itself that is special, but rather the operatorial realization within the worldline formalism.

The action of the theory can be computed in the worldline formalism following the same strategy as in the previous section. Skipping the details, we report here a brief outline. The only differences are that $c$ is replaced by $\mathfrak{c}$ in the definition \eqref{InnerProductI} of the inner product, and that the operator $\alpha_{\mu\nu}$ in the basis for the Hilbert space is now replaced by the combination $\alpha_\mu\,\alpha_\nu$ of the vector operator $\alpha_\mu$. Namely, the BRST charge is Hermitian if the analogous conditions of \eqref{HermitianCond1} and \eqref{HermitianCond2} hold
\begin{equation}\label{HermitianCondition}
\alpha_\mu^\dagger = \bar\alpha_\mu = \frac{\delta}{\delta \alpha^\mu}, \quad
\mathfrak{c}^\dagger = \mathfrak{c}, \quad
C^\dagger = \bar{C} = \frac{\delta_L}{\delta C} \;\;\Rightarrow  \;\;
\mathfrak{b}^\dagger = \mathfrak{b} = \frac{\delta_L}{\delta\mathfrak{c}}, \quad
B^\dagger = \bar{B} = \frac{\delta_L}{\delta B},
\end{equation}
and the unique non-vanishing inner product involving $\alpha_\mu\,\alpha_\nu$ are
\begin{equation}
\braket{\alpha_\mu\,\alpha_\nu,\alpha^\vrho\,\alpha^\sigma\,c} = \braket{\alpha^\vrho\,\alpha^\sigma\,c,\alpha_\mu\,\alpha_\nu}= i\,\delta_{(\mu}^\vrho\,\delta_{\nu)}^\sigma.
\end{equation}
The BV action of the fracton theory described by the worldline model under discussion turns out to proportional to the spacetime integral of the inner product between $\hat{\psi}$ and its BRST variation, similarly to the previous section, again requiring all the fields to be real, except for the ghost, which is assumed to be pure imaginary:
\begin{equation}
\Gamma_{\textsc{bv}}^{\text{(fr)}}(\alpha = -\tfrac{3}{8},\beta = \tfrac{1}{4}) = -\frac{i}{4}\int\diff^D x\,\braket{\hat{\psi},\mathcal{Q}\,\hat{\psi}}.
\end{equation}

\subsection{Deformed vector model}\label{modelIII}

In this section, a deformation of the model in the previous section is investigated, in order to reproduce more general covariant fracton gauge theories. Even if the spacetime theory we want to reproduce is free, we can employ the same techniques originally introduced in dealing with interacting worldline models. The latter are usually realized by deforming a free worldline model, by taking into account possible minimal and non-minimal couplings with background fields. Here, there are no background fields, but the BRST cohomology can be studied along the same lines \cite{Dai:2008bh, Bonezzi:2018box, Bonezzi:2020jjq, Fecit:2023kah, Bastianelli:2025khx}.

Consider again the following operators:
\begin{equation}
L=(\alpha^{\mu} p_\mu)^2\;, \quad \bar L= (\bar\alpha^{\mu} p_\mu)^2\;, \quad J=\tfrac12 \alpha^{\mu}\bar \alpha_{\mu}.
\end{equation} 
Keeping them as they stand is reasonable, since, as discussed in the Introduction, we need a double-momenta constraint as $L$ in order to reproduce the BRST symmetry $\sim \partial_\mu\,\partial_\nu \,\lambda$ of covariant fracton theories, and we do not want to introduce prefactors in such a transformation, so that $L$ should not be deformed. Therefore, also $\bar L$ remains undeformed. Thus, it remains to consider the operator $\ell_{\mathrm{w}}$ in \eqref{SymmetricEll}. The simplest way to deform it is to introduce a pair of free parameters $(\tilde \alpha,\tilde \beta)$, in such a way that, when, say, they are set to zero, the undeformed operator \eqref{SymmetricEll} is recovered:\footnote{In the definition \eqref{deformed}, the quantum operators are to be understood as acting in this precise order. Since we want to interpret this model as a deformation of the one of Section~\ref{modelII}, we assume that $\ell_{\mathrm{w}}(\tilde\alpha,\tilde\beta)$ corresponds to the Weyl-ordered quantized version of some classical phase-space function to be later identified.}
\begin{align}\label{deformed}
\ell_{\mathrm{w}}(\tilde\alpha,\tilde\beta) = (1+\tilde\beta)\, \alpha^\mu\,p_\mu\,\bar\alpha^\nu\,p_\nu+(\tfrac{1}{2}+\tilde\alpha)\, H,\;\;\text{such that}\;\;\ell_{\mathrm{w}}(0,0) = \ell_{\mathrm{w}}.
\end{align}
As a consequence, the algebra of the constraints becomes
\begin{equation}\label{algebraconstraintsIII@Q}
[\ell_{\mathrm{w}}(\tilde\alpha,\tilde\beta),L]=2\,(1+\tilde\beta)\,H\,L, \quad [\ell_{\mathrm{w}}(\tilde\alpha,\tilde\beta),\bar L]=-2\,(1+\tilde\beta)\,H\,\bar L,
\end{equation}
\begin{equation}\label{algebraconstraintsIII@Qbis}
[\bar{L},L] = \tfrac{4}{1+\tilde{\beta}}\,H\,(\ell_{\mathrm{w}}(\tilde\alpha,\tilde\beta)-\tfrac{2\,\tilde\alpha - \tilde\beta}{2}\,H), \quad
\end{equation}
while the commutators \eqref{algebraconstraintsII@Q.2} are unaltered (with $\ell_{\mathrm{w}}$ replaced by $\ell_{\mathrm{w}}(\tilde\alpha,\tilde\beta)$). Once the constraints and their algebra are established, one can enlarge the Hilbert space by adding pairs of antighost-ghost for each constraint, as in \eqref{HilbertModelII}, and one can define the nilpotent BRST charge, starting from  
\begin{equation}\label{min}
\mathcal{Q} = \mathfrak{c}\,\ell_{\mathrm{w}}(\tilde\alpha,\tilde\beta) + C\,\bar L+ \bar C\, L +\mathcal{M},
\end{equation}
where again a suitable $\mathcal{M}$ must be identified in order to make $\mathcal{Q}$ such that $\mathcal{Q}^2 = 0$. We try to guess:
\begin{equation}\label{ansatz}
\mathcal{Q} =\mathfrak{c}\,\ell_{\mathrm{w}}(\tilde\alpha,\tilde\beta)+C\,\bar L+\bar C\, L+\kappa_1\,C\,\bar C\,\mathfrak{b}\,H +\kappa_2\,\bar C\, B\,\mathfrak{c}\,H +\kappa_3\,C\,\bar B\,\mathfrak{c}\,H,
\end{equation}
for arbitrary constants $\kappa_1, \kappa_2, \kappa_3$. This ansatz can be justified by looking at the quantum algebra \eqref{algebraconstraintsIII@Q} and trying to come up with a BRST operator using the aforementioned standard techniques \cite{VanHolten:2001nj}, together with recalling that it should reduce to the undeformed BRST charge \eqref{QII} when $\tilde\alpha$ and $\tilde\beta$ are sent to zero.

In interacting worldline models \cite{Dai:2008bh, Bonezzi:2018box, Fecit:2023kah, Bastianelli:2025khx}, the BRST charge is deformed to accommodate possible couplings with background fields, and the latter are then constrained by the nilpotency requirement. It is often the case that nilpotency cannot be achieved in the full BRST Hilbert space, but only on a subspace, identified with an eigenspace of a suitable ghost-extended number operator $\mathcal{J}$. This is consistent whenever $\mathcal{J}$ and $\mathcal{Q}$ commute, in such a way that the eigenspaces of $\mathcal{J}$ are stable under the action of $\mathcal{Q}$. Eventually, the consistency of this procedure should also be tested by constructing the path integral for such models.

Even if the expected spacetime theory is free, we follow the same strategy by considering the ghost-extended number operator of Section~\ref{modelII}
\begin{equation}
\mathcal{J}=\frac{1}{2}\,\alpha^{\mu}\,\bar\alpha_{\mu} + C \,\bar B + B\,\bar C+\frac{D}{4},
\end{equation}
which commutes with $\mathcal{Q}$ for any choice of the deformation parameters $\tilde\alpha,\tilde\beta$, and of the constant $\kappa_1,\kappa_2,\kappa_3$. It is then consistent to require that $\mathcal{Q}$ is nilpotent on the $\mathcal{J}$ eigenspace with eigenvalue $\tfrac{D+4}{4}$, i.e.
\begin{equation}\label{Q2}
\mathcal{Q}^2\,\psi = 0, \quad \forall\;\psi\;\text{such that}\;J\,\psi= \tfrac{D+4}{4}\,\psi.
\end{equation}
This requirement uniquely fixes the constants $\kappa_1,\kappa_2$ and $\kappa_3$:
\begin{equation}
\kappa_1 = -\frac{4}{1+2\,\tilde\alpha}, \quad
\kappa_3 = -\kappa_2 = 2\,(1+\tilde\beta),
\end{equation}
so that, replacing in \eqref{ansatz}, assuming that $\tilde\alpha \neq -\frac{1}{2}$, one gets
\begin{align}
\mathcal{Q} &=\mathfrak{c}\,\ell_{\mathrm{w}}(\tilde\alpha,\tilde\beta)+C\,\bar L+\bar C\, L-\tfrac{4}{1+2\,\tilde\alpha}\,C\,\bar C\,\mathfrak{b}\,H +2\,(1+\tilde\beta)(\,C\,\bar B-\bar C\, B)\,\mathfrak{c}\,H  \nonumber \\
&= (\tfrac{3}{2} -\tilde\alpha  + 2\,\tilde\beta)\,\mathfrak{c}\,\de^2 
-(1+\tilde\beta)\,\mathfrak{c}\,\alpha^\mu\,\alpha^\nu\,\de_\mu\,\de_\nu 
-C\,\tfrac{\delta_L}{\delta\alpha_\mu}\,\tfrac{\delta_L}{\delta\,\alpha_\nu}\,\de_\mu\,\de_\nu - \bar{C}\,\alpha^\mu\,\alpha^\nu\,\de_\mu\,\de_\nu \nonumber\\
&\phantom{=} + \tfrac{4}{1+2\,\tilde\alpha}\,C\,\tfrac{\delta_L}{\delta B}\,\tfrac{\delta_L}{\delta \mathfrak{c}}\,\de^2 - 2\,(1+\tilde\beta)\,\mathfrak{c}\,(C\,\tfrac{\delta_L}{\delta C}+B\,\tfrac{\delta_L}{\delta B})\,\de^2.
\end{align}

The most general state $\psi$ in the $\mathcal{J}$ eigenspace with eigenvalue $\tfrac{D+4}{4}$ is the same as in the undeformed model \eqref{psinottracedII}, with the same choice of parity and ghost number. The suitable ``trace operator" for reducing the components, namely, which should satisfy
\begin{equation}
[\mathcal{T},\mathcal{J}]=\mathcal{T}, \quad [\mathcal{T},Q]\,\psi = 0, \quad \forall\;\psi\;\text{such that}\;J\,\psi=\tfrac{D+4}{4}\,\psi,
\end{equation}
is deformed in the following way:
\begin{align}
\mathcal{T}= \frac{1}{2}\,\bar\alpha^{\mu}\,\bar\alpha_{\mu}+\frac{2}{1+2\,\tilde\alpha}\,\mathfrak{b}\,\bar C-(1+\tilde \beta)\,\mathfrak{c}\,\bar B,
\end{align}
which consistently is equal to \eqref{TraceII} when $\tilde\alpha,\tilde\beta \rightarrow 0$. The constraints on $\psi$ imposed by the requirement that it sits in the kernel of $\mathcal{T}$ are
\begin{align}\label{TpsiIII}
\mathcal{T}\,\psi = 0 \Leftrightarrow 
\phi =-\frac{1+2\tilde \alpha}{2}\,h, \; \tilde{\phi} =\frac{1}{1+\tilde \beta}\,\tilde{h}.
\end{align}
Therefore, if we replace in \eqref{psinottracedII}, assuming that $\tilde\beta \neq -1$, we get
\begin{align}\label{stringfieldIII}
\hat{\psi}= h_{\mu\nu}\,\alpha^{\mu}\,\alpha^{\nu}
+\lambda\,B 
+\tfrac{1}{1+\tilde \beta}\,\tilde{h}\,C 
-\tfrac{1+2\tilde \alpha}{2}\,h\,B\,\mathfrak{c} 
+\tilde{\lambda}\,C\,\mathfrak{c} 
+\tilde{h}_{\mu\nu}\,\alpha^{\mu}\alpha^{\nu}\,\mathfrak{c},\;\;\text{such that}\;\mathcal{T}\,\hat\psi = 0.
\end{align}
Now, we can compute the deformed BRST transformations on the field components, always according to the definition of the BRST differential operator $s$ in \eqref{BRSTdifferential}. The result is
\begin{equations}
s\,\lambda &=0,\label{s1}\\    
s\,h_{\mu\nu} &= \partial_\mu\,\partial_\nu\,\lambda,\label{s2}\\
s\,h&=\partial^2 \lambda,\label{s3}\\
s\,\tilde{h}_{\mu\nu} &= (\tfrac{3}{2}-\tilde\alpha+ 2\,\tilde \beta)\,\partial^2 h_{\mu\nu}-(1+\tilde\beta)\,\partial^\vrho\,\partial_{(\mu}\,h_{\nu)\vrho}+(\tfrac{1}{2}+\tilde \alpha)\,\partial_\mu\,\partial_\nu\,h,\label{s4}\\
s\,\tilde{h} &= 2\,(1+\tilde \beta)\,(\partial^2 h- \partial_\mu \partial_\nu\,h^{\mu\nu}),\label{s5}\\   
s\,\tilde{\lambda} &= -\tfrac{1+2\tilde \alpha}{2+2\tilde \beta}\,\partial^2\,\tilde{h}+2\,\partial_\mu\,\partial_\nu\,\tilde{h}^{\mu\nu}.\label{s6}
\end{equations}
Consistently, setting $\tilde\alpha,\tilde\beta \rightarrow 0$, we recover to the undeformed transformations \eqref{eomsII1}--\eqref{eomsII2}. The transformations \eqref{s3} and \eqref{s5} are redundant, since they are the trace of \eqref{s2} and \eqref{s4} respectively. The transformation \eqref{s1} and \eqref{s2}, which are left undeformed, are the familiar ones of covariant fracton gauge theories. In order to compare with the BRST transformations on the antifields sector in \eqref{BRSTFractonTot3}--\eqref{BRSTFractonTot4}, consider the following redefinition of the deformation parameters\footnote{Since $\tilde\alpha = -\frac{1}{2}$ and $\tilde\beta = -1$ are excluded, and $\tilde \alpha=-\frac{1}{2} + 2\,\beta$, $\tilde \beta=-1-2\,\alpha+\beta$, the cases $2\,\alpha  - \beta = 0$ and $\beta = 0$ are correspondingly excluded.}
\begin{align}\label{RedefinitionDeformation}
\alpha = -\tfrac{3}{8} + \tfrac{\tilde\alpha}{2}-\tilde\beta, \quad
\beta = \tfrac{1}{4} + \tfrac{\tilde\alpha}{2},
\end{align}
and the following change of variables 
\begin{equation}
h^*_{\mu\nu} = \tilde{h}_{\mu\nu}+\tfrac{\beta}{2\,\alpha-\beta}\,\eta_{\mu\nu}\,\tilde{h}, \quad \lambda^* = \tfrac{1}{2}\,\tilde{\lambda},
\end{equation}
which is consistent when $2\,\alpha - \beta \neq 0$ and $2\,\alpha + (D-1)\,\beta\neq 0$.\footnote{Indeed, the inverse rules are: $\tilde{h}_{\mu\nu} = h^*_{\mu\nu} - \frac{\beta}{2\,\alpha + (D-1)\,\beta}\,\eta_{\mu\nu}\,h^*$, $\tilde{\lambda} = 2\,\lambda^*$.} Replacing in \eqref{s1}--\eqref{s6}, we obtain the BRST transformations in \eqref{BRSTFractonTot1}--\eqref{BRSTFractonTot4}, if we identify the parameters $\alpha$ and $\beta$ in \eqref{RedefinitionDeformation} with the constant identifying the covariant fracton gauge theories. Indeed, if $\tilde\alpha,\tilde\beta \rightarrow 0$ in \eqref{RedefinitionDeformation} we obtain consistently the choice of the constants in \eqref{ConstModelI}.

Finally, we conclude that the deformed vector model reproduces all the covariant fracton gauge theories as spacetime theories, except in the cases in which
$\beta = 0$, $2\,\alpha - \beta = 0$, $2\,\alpha + (D-1)\,\beta = 0$. Remember that the peculiar case in which $2\,\alpha - \beta = 0$ is captured by the model discussed in Section~\ref{modelI}, and that $2\,\alpha + (D-1)\,\beta = 0$ is the traceless limit (see Eq.~\eqref{TracelessCond}).

One can verify that, using \eqref{HermitianCondition}, the deformed BRST charge is hermitian with respect to the inner product \eqref{InnerProduct}. As in the undeformed case, one can show that the BV action can be reproduced by compute the spacetime integral of the inner product between the BRST variation of $\hat{\psi}$ and $\hat{\psi}$ itself, again assuming that the ghost $\lambda$ is pure imaginary and the other fields real:
\begin{equation}
\Gamma_{\textsc{bv}}^{\text{(fr)}}(\alpha,\beta) = -\frac{i}{2}\int\diff^D x\,\braket{\hat\psi,\mathcal{Q}\,\hat\psi} = \int\diff^D x\,(\alpha\,f^{\mu\nu\vrho}\,f_{\mu\nu\vrho} + \beta\,f^{\mu\nu}{}_\nu\,f_{\mu\vrho}{}^\vrho + h^{*\mu\nu}\,\de_\mu\,\de_\nu\,\lambda),
\end{equation}
for the allowed values of $\alpha$ and $\beta$.

While having a BRST system that correctly reproduces the desired results is certainly valuable, it remains essential to identify the underlying classical worldline theory, which, upon quantization, gives rise to this BRST structure. As already outlined at the beginning of this section, we do this by regarding the present model as a deformation of the vector model. Thus, the phase space is defined by the relations \eqref{algebraphspII} and the following phase-space functions
\begin{equation}
L=(\alpha^{\mu} p_\mu)^2, \quad \bar L= (\bar\alpha^{\mu} p_\mu)^2, \quad J=\tfrac12 \alpha^{\mu}\bar \alpha_{\mu}
\end{equation} 
should still be implemented to constrain the dynamics of the worldline model, as they stand. On the other hand, the remaining classical constraint should take the form
\begin{equation}\label{deformedclassical}
\ell(\tilde\alpha,\tilde\beta) =(1+\tilde \beta)\, \alpha^\mu\,p_\mu\,\bar\alpha^\nu\, p_\nu+(\tilde\alpha-\tfrac{\tilde\beta}{2})\, H,
\end{equation}
since it is converted into the quantum constraint \eqref{deformed} upon quantization by choosing the symmetric prescription, just as chosen in the previous section. As a consistency check, note that the set of classical constraints reduces to that of the vector model, as in \eqref{constraintsII}, if we set the deformation parameters to zero. The Poisson brackets of the constraints and of the number operator are:
\begin{equation}
\{\ell(\tilde\alpha,\tilde\beta) ,L\}=-2\,i\,(1+\tilde \beta)H\,L,\quad
\{\ell(\tilde\alpha,\tilde\beta),\bar L\}=2\,i\,(1+\tilde\beta)\,H\,\bar L,
\end{equation}
\begin{equation}
\{\bar L,L\} =-\tfrac{4\,i}{1+\tilde \beta}\,H\,(\ell(\tilde\alpha,\tilde\beta) -(\tilde \alpha-\tfrac{\tilde \beta}{2}) \,H), \quad
\{J,L\}=-i\,L, \quad \{J,\bar L\}=i\,\bar L,
\end{equation}
and the worldline action is
\begin{equation}
S = \int \diff\tau\,(p_\mu\,\dot x^\mu -i\,\bar\alpha^\mu\, \dot\alpha_{\mu}- \omega\,\ell(\tilde\alpha,\tilde\beta)- u\,\bar{L} - \bar{u}\,L - a\,J),
\end{equation}
with the usual set of gauge fields $(\omega,u,\bar u,a)$. This action is not left invariant under the gauge transformations generated by $G=\chi \,\ell + \varepsilon\,\bar{L} + \bar{\varepsilon}\,L + \phi\,J$ via Poisson brackets with phase-space variables and gauge fields. The same happens in interacting worldline models, where the BRST charge on a specific eigenspace of the ghost-extended number operator is nilpotent, while the underlying classical gauge symmetry is generally broken. Interestingly, the same situation is met here even if the spacetime theory described by the deformed vector model is free.

\section{Gauge-fixing}\label{gauge-fixing}

\subsection{BV-BRST gauge-fixing}

In this section, we study the gauge-fixing in covariant fracton theories from the spacetime. In the next one, we will study the same problem from the worldline perspective, and we make a comparison.

By gauge-fixing one means the inclusion of a Lorentz-invariant term in the action, which breaks completely or partially the gauge symmetry, in such a way that the kinetic operator in momentum space becomes invertible. In BV formalism, the gauge-fixing procedure consists in defining an anticommuting functional $\psi[\vphi]$ of the fields $\vphi^i$ of the theory, with ghost number $g_\psi = -1$, and in imposing the antifields $\vphi_i^*$ to be equal to the variation of $\psi[\vphi]$ with respect to $\vphi^i$ \cite{Fuster:2005eg}:
\begin{equation}\label{GaugeFixingAntiFields}
\vphi_i^* = \frac{\delta\,\psi[\vphi]}{\delta\,\vphi^i}, \;\;\forall\;\text{field}\;\vphi\;\text{of the theory.}
\end{equation}
It is not necessary to specify if the functional derivative is right or left, since, if $\vphi^i$ is anticommuting, then the coefficient of $\vphi^i$ in $\psi[\vphi]$ must be commuting, since $\psi[\vphi]$ is required to be anticommuting. The gauge-fixed action $\Gamma_\psi[\vphi] $ with respect to $\psi[\vphi]$ is obtained by replacing the condition \eqref{GaugeFixingAntiFields} in the BV action:
\begin{equation}
\Gamma_\psi[\vphi] = \Gamma_{\textsc{bv}}[\vphi,\,{\vphi^* = \delta\,\psi/\delta\,\vphi}].
\end{equation}
Replacing the condition \eqref{GaugeFixingAntiFields} in the BRST transformations of the antifields one gets the equations of motion of the gauge-fixed action as consistency conditions.

In the case of covariant fracton, we have two possible covariant gauge fixing conditions, either a combination of the double divergence and of the Laplacian of the trace of the fracton gauge field, or the trace itself to be set to zero:
\begin{equation}\label{GaugeFixingFracton}
\de_\mu\,\de_\nu\,h^{\mu\nu} + k\,\de^2\,h \overset{!}{=} 0 \quad \text{or} \quad h \overset{!}{=} 0,
\end{equation}
where $k \neq -1$ is a constant.\footnote{The case $k=-1$ is excluded because in that case the combination is gauge invariant: $\de_\mu\,\de_\nu\,h^{\mu\nu}-\de^2\,h = -\de_\mu\,f^{\mu\nu}{}_\nu$.}
In order to impose one of the two conditions, we enlarge the space of fields of the theory by introducing a trivial doublet $(\bar{\lambda},\chi)$ 
\begin{equation}\label{BRSTDoublet} 
s\,\bar\lambda = \chi, \quad 
s\,\chi = 0,
\end{equation}
which does not change the BRST cohomology \cite{Brandt:1989rd}, \cite{Piguet:1995er}. $\bar{\lambda}$ is usually called ``antighost", but it is not to be confused with antifields; $\chi$ plays the role of a Lagrangian multiplier for the gauge fixing condition. Requiring $\chi$ to be an even scalar field with vanishing ghost number, the BRST rule \eqref{BRSTDoublet} imposes $\bar\lambda$ to be an odd scalar field with ghost number $-1$:
\begin{equation}
|\chi| = 0, \quad |\bar\lambda| = 1, \quad 
\text{gh}(\chi) = 0, \quad \text{gh}(\bar\lambda) = -1.
\end{equation}
The doublet contributes to the BV action through 
\begin{equation}
\Gamma_{\textsc{bv}} = \Gamma_{\text{bv}}^{\text{(fr)}} - \bar\lambda^*\,s\,\bar\lambda\, - \chi^*\,s\,\chi\, = \Gamma_{\text{bv}}^{\text{(fr)}} - \bar\lambda^*\,\chi,
\end{equation}
where $\bar\lambda^*$ and $\chi^*$ are the antifields of $\bar\lambda $ and $\chi$ respectively. They are such that
\begin{equation}
|\chi^*| = 1, \quad
|\bar\lambda^*| = 0, \quad
\text{gh}(\chi^*) = -1, \quad
\text{gh}(\bar\lambda^*) = 0.
\end{equation}
Then, using \eqref{sphisphistar}, the BRST rules of the antifields of the doublet are
\begin{equation}
s\,\chi^* = -\bar\lambda^*, \quad s\,\bar\lambda^* = 0.
\end{equation}
Let us introduce the following two gauge-fixing functionals, corresponding to the two gauge-fixing conditions in \eqref{GaugeFixingFracton},
\begin{equation}\label{GaugeFixingFunctional}
\psi_1 = \int\diff^D x\,\bar\lambda\,(\de_\mu\,\de_\nu\,h^{\mu\nu}+ k\,\de^2\,h) \quad\text{or}\quad
\psi_2 = \int\diff^D x\,\bar\lambda\,h.
\end{equation}
Computing their variation with respect to the fields of the theory, we obtain the expressions for the antifields, according to \eqref{GaugeFixingAntiFields}. In the first case,
\begin{equation}
\lambda^* = \frac{\delta \psi_1}{\delta \lambda} = 0, \;\;
h^*_{\mu\nu} = \frac{\delta \psi_1}{\delta h^{\mu\nu}} = \de_\mu\,\de_\nu\,\bar\lambda + k\,\de^2\,\bar\lambda\,\eta_{\mu\nu}, \;\;
\bar\lambda^* = \frac{\delta \psi_1}{\delta \bar\lambda} = \de_\mu\,\de_\nu\,h^{\mu\nu} + k\,\de^2\,h, \;\;
\chi^* = \frac{\delta \psi_1}{\delta \chi} = 0.
\end{equation}
Instead, in the second one,
\begin{equation}
\lambda^* = \frac{\delta \psi_2}{\delta \lambda} = 0, \quad
h^*_{\mu\nu} = \frac{\delta \psi_2}{\delta h^{\mu\nu}} = \eta_{\mu\nu}\,\bar\lambda, \quad
\bar\lambda^* = \frac{\delta \psi_2}{\delta \bar\lambda} = h, \quad
\chi^* = \frac{\delta \psi_2}{\delta \chi} = 0.
\end{equation}
Finally, the gauge-fixed action is 
\begin{align}\label{GaugeFixedActionFractonI}
\Gamma^{\text{(fr)}}_{\psi_1} &= S^{\text{(fr)}}  + \int\diff^D x\,[(1+k)\,\bar\lambda\,(\de^2)^2\,\lambda - \chi\,(\de_\mu\,\de_\nu\,h^{\mu\nu}+k\,\de^2\,h)]\nonumber \\
& = \int\diff^D x\,[-\,2\,\alpha\,h^{\mu\nu}\,\de^2\,h_{\mu\nu} -\beta\,h\,\de^2\,h- (2\,\alpha-\beta)\,\de_\mu\,h^{\mu\nu}\,\de^\vrho\,h_{\vrho\nu} \nonumber\\
& \qquad\qquad\;\; + 2\,\alpha\,h\,\de_\mu\,\de_\nu\,h^{\mu\nu} + (1+k)\,\bar\lambda\,(\de^2)^2\,\lambda - \chi\,(\de_\mu\,\de_\nu\,h^{\mu\nu}+k\,\de^2\,h)],
\end{align}
in the first case, and
\begin{align}\label{GaugeFixedActionFractonII}
\Gamma^{\text{(fr)}}_{\psi_2} &= S^{\text{(fr)}} + \int\diff^D x\,(\bar\lambda\,\de^2\,\lambda - \chi\,h)  \nonumber \\
&= \int\diff^D x\,(-\,2\,\alpha\,h^{\mu\nu}\,\de^2\,h_{\mu\nu} -\beta\,h\,\de^2\,h- (2\,\alpha-\beta)\,\de_\mu\,h^{\mu\nu}\,\de^\vrho\,h_{\vrho\nu} \nonumber\\
& \qquad\qquad\;\; + 2\,\alpha\,h\,\de_\mu\,\de_\nu\,h^{\mu\nu} + \bar\lambda\,\de^2\,\lambda - \chi\,h),
\end{align}
in the second one.\footnote{Notice that the gauge-fixing procedure can be realized directly in BRST formalism without using  antifields, by adding to the physical action the BRST of the gauge-fixing functional \eqref{GaugeFixingFunctional}: $\Gamma^{\text{(fr)}}_{\psi_i} = S^{\text{(fr)}}-s\,\psi_i$.}

The equation of motion of the Lagrangian multiplier $\chi$ corresponds to the gauge fixing condition. The equations of motion of $h_{\mu\nu}$ are 
\begin{align}
\frac{\delta\Gamma_{\psi_1}^{\text{(fr)}}}{\delta h^{\mu\nu}} &= -\,4\,\alpha\,\de^2\,h_{\mu\nu} + (2\,\alpha - \beta)\,\de^\vrho\,\de_{(\mu}\,h_{\nu)\vrho} + 2\,\beta\,\de_\mu\,\de_\nu\,h \nonumber\\
& \phantom{=} +  2\,\beta\,\eta_{\mu\nu}\,(\de_\vrho\,\de_\sigma\,h^{\vrho\sigma} - \de^2\,h) + (\de_\mu\,\de_\nu + k\,\eta_{\mu\nu}\,\de^2)\,\chi,
\end{align}
whose trace is
\begin{equation}
\eta^{\mu\nu}\,\frac{\delta\Gamma_{\psi_1}^{\text{(fr)}}}{\delta h^{\mu\nu}} = 2\,(2\,\alpha + (d-1)\,\beta)(\de_\mu\,\de_\nu\,h^{\mu\nu} - \de^2\,h) + (1+k\,d)\,\de^2\,\chi.
\end{equation}
Notice that $\chi$ cannot be algebraically eliminated, whatever the value of $k$. Instead, this is possible for the second gauge fixing $h=0$. Indeed, the equations of motion are
\begin{align}\label{EomGaugeFix2}
\frac{\delta\Gamma_{\psi_2}^{\text{(fr)}}}{\delta h^{\mu\nu}} &= -\,4\,\alpha\,\de^2\,h_{\mu\nu} + (2\,\alpha - \beta)\,\de^\vrho\,\de_{(\mu}\,h_{\nu)\vrho} + 2\,\beta\,\de_\mu\,\de_\nu\,h \nonumber\\
& \phantom{=} +  2\,\beta\,\eta_{\mu\nu}\,(\de_\vrho\,\de_\sigma\,h^{\vrho\sigma} - \de^2\,h) + \eta_{\mu\nu}\,\chi,
\end{align}
so that, taking the trace, 
\begin{equation}
\chi = -\tfrac{2}{D}\,(2\,\alpha + (D-1)\,\beta)(\de_\mu\,\de_\nu\,h^{\mu\nu} - \de^2\,h).
\end{equation}
Replacing in \eqref{EomGaugeFix2}, one gets
\begin{equation}\label{GaugeFixedPart2}
-4\,\alpha\,\de^2\,h_{\mu\nu} + (2\,\alpha - \beta)\,\de^\vrho\,\de_{(\mu}\,h_{\nu)\vrho} + 2\,\beta\,\de_\mu\,\de_\nu\,h - \tfrac{2}{D}\,(2\,\alpha - \beta)\,\eta_{\mu\nu}\,(\de_\vrho\,\de_\sigma\,h^{\vrho\sigma} - \de^2\,h) = 0,
\end{equation}
which is traceless.

\subsection{Siegel gauge}

Let us examine now the outcome of the gauge-fixing procedure from the worldline perspective. The so-called \emph{Siegel gauge} is implemented by requiring that $\psi$ does not depend on the ``reparametrization ghost", responsible for the doubling of the BRST Hilbert space basis: this is $c$ in the tensor model, and $\mathfrak{c}$ in the vector model and in its deformation.\footnote{Actually in the vector models, strictly speaking, there is no reparametrization ghost because the Hamiltonian is not directly treated as a constraint, but the first-quantization ghost $\mathfrak{c}$ play a similar role as $c$ in the model in Section~ref{modelI}: namely, it doubles the BRST Hilbert space and it is integrated in the inner product.} In other words, for the tensor model, $\psi$ is required to sit in the kernel of $b$
\begin{equation}\label{SiegelI}
b\,\psi = \frac{\delta_L}{\delta c}\,\psi \overset{!}{=} 0.
\end{equation}
Here and in the next equations we use the operator $b$ of the tensor model, but the similar equations hold for the vector models with $\mathfrak{b} = \frac{\delta_L}{\delta\mathfrak{c}}$ in place of $b$. Crucially, we implement the Siegel gauge on $\psi$, which lives in the sector of the BRST Hilbert space with $\mathcal{J}=1$, and not on $\hat{\psi}$, which is also restricted in the subsector with $\mathcal{T} = 1$ -- otherwise, we would obtain too restrictive conditions. The gauge fixed action for the Siegel gauge is given by \cite{Siegel:1984xd, Asano:2006hk}
\begin{equation}
-\frac{i}{2}\int\diff^D x\left(\braket{\psi,Q\,\psi}+\braket{\zeta,b\,\psi}\right),
\end{equation}
where $\zeta$ is an even state with ghost number two, which plays the role of the Nakanishi-Lautrup field, since its equation of motion corresponds to the Siegel gauge condition \eqref{SiegelI}. Replacing this condition back in the action, one gets a partially on-shell action
\begin{equation}
S_{\text{Siegel}} = -\frac{i}{2}\int\diff^D x\,\braket{\tilde\psi,Q\,\tilde\psi},
\end{equation} 
where $\tilde\psi$ denotes the arbitrary state satisfying \eqref{SiegelI}. Using the definition of the inner product \eqref{InnerProductI}, one can perform the integration over the fermionic variable:
\begin{align}
S_{\text{Siegel}} &= -\frac{i}{2}\int\diff^D x\,\braket{\tilde\psi,Q\,\tilde\psi} = \frac{1}{2}\int \diff^D x\int\diff c\,\overline{\tilde\psi}\,Q\,\tilde\psi = \frac{1}{2}\int\diff^D x\,\frac{\delta_L}{\delta c}\,(\overline{\tilde\psi}\,Q\,\tilde\psi) \nonumber \\
&= \frac{1}{2}\int\diff^D x\,\overline{\tilde\psi}\,\frac{\delta_L}{\delta c}\,Q\,\tilde\psi = \frac{1}{2}\int\diff^D x\,\overline{\tilde\psi}\,\left[\frac{\delta_L}{\delta c},Q\right]\,\tilde\psi  = \frac{1}{2}\int\diff^D x\,\overline{\tilde\psi}\,[b,Q]\,\tilde\psi.
\end{align} 
Therefore, the equations of motion are encoded in 
\begin{equation}
[b,Q]\,\tilde\psi = 0.
\end{equation}
In the tensor model the commutator $[b,Q]$ is simply the Hamiltonian 
\begin{equation}
[b,Q] = H = -\de^2.
\end{equation}
Thus, the action gauge-fixed action is a diagonal quadratic form, and the corresponding equations of motion are wave equations. But in principle, it is not necessary for a gauge-fixed action to be diagonal, and one could expect non-diagonal actions in a more general case. This is precisely what happens in the vector models, where the dependence on $\mathfrak{c}$ is more complicated than the one on $c$ in the tensor model. In the deformed case commutator reads
\begin{align}
[\mathfrak{b},Q] &= \ell_{\mathrm{w}}(\alpha,\beta) -2\,(2\,\alpha-\beta)\,(C\,\bar{B}-\bar{C}\,B)\,H  \nonumber\\
&= -4\,\alpha\,\de^2 +(2\,\alpha-\beta)\,\alpha^\mu\frac{\delta_L}{\delta\alpha_\nu}\,\de_\mu\,\de_\nu + 2\,(2\,\alpha - \beta)\,(B\,\tfrac{\delta_L}{\delta B} + C\,\tfrac{\delta_L}{\delta C})\,\de^2,
\end{align}
where $\ell_{\mathrm{w}}(\alpha,\beta) \equiv \ell_{\mathrm{w}}(\tilde\alpha(\alpha,\beta),\tilde\beta(\alpha,\beta))$. The undeformed case is recovered setting $(\alpha,\beta) \rightarrow (-\frac{3}{8},\frac{1}{4})$.

Consider the tensor model. The Siegel-gauge condition \eqref{SiegelI} imposes the following constraints on the field components of $\psi$:
\begin{equation}
\phi = 0, \quad \hat{\tilde{h}}_{\mu\nu}= 0, \quad \lambda^* = 0,
\end{equation}
so that we consider the following reduced state
\begin{equation}
\tilde\psi = h_{\mu\nu}\,\alpha^{\mu\nu} + \lambda\,B + \tilde{\phi}\,C, \;\;\text{such that}\;b\,\tilde\psi = 0,
\end{equation}
and we can compute the gauge-fixed action, according to 
\begin{equation}\label{GaugeFixedActionFractonModelIWorldline}
S_{\text{Siegel}} = \frac{1}{2}\int\diff^D x\,\overline{\tilde\psi}\,H\,\tilde\psi = -\frac{1}{2}\,\int\diff^D x\,(h^{\mu\nu}\,\de^2\,h_{\mu\nu} + \tilde{\phi}\,\de^2\,\lambda).
\end{equation}
The equations of motion are the wave equation for the three field components in $\tilde\psi$:
\begin{equation}\label{EomModelIGaugeFixed}
\de^2\,h_{\mu\nu} = 0, \quad \de^2\,\tilde{\phi} = 0, \quad \de^2\,\lambda = 0.
\end{equation}
Restricting to the physical sector (at ghost number zero), 
the first equation can be obtained by varying the partially on-shell BV-BRST gauge-fixed action, obtained by imposing the gauge fixing condition in the action \eqref{GaugeFixedActionFractonI} (in the case $k=0$, where the gauge fixing condition is $\de_\mu\,\de_\nu\,h^{\mu\nu} = 0$), before varying with respect to the gauge fixing:
\begin{equation}
{\Gamma_{\psi_1}^{\text{(fr)}}}_{|_{\de_\mu\de_\nu h^{\mu\nu}\rightarrow 0}} = \int \diff^D x\,[-2\,\alpha\,h^{\mu\nu}\,\de^2\,h_{\mu\nu} -(2\,\alpha-\beta)\,\de_\mu\,h^{\mu\nu}\,\de_\vrho\,h_\nu{}^\vrho -\beta\,h\,\de^2\,h + \,\bar\lambda\,(\de^2)^2\,\lambda].
\end{equation}
The equations of motion of $h_{\mu\nu}$ are
\begin{equation}\label{EomsPartialGaugeFixed}
-4\,\alpha\,\de^2\,h_{\mu\nu} + (2\,\alpha - \beta)\,\de^\vrho\,\de_{(\mu}\,h_{\nu)\vrho} - 2\,\beta\,\eta_{\mu\nu}\,\de^2\,h = 0,
\end{equation}
whose trace, upon using $\de_\mu\,\de_\nu\,h^{\mu\nu} = 0$, is
\begin{equation}
-2\,(2\,\alpha + \beta\,D)\,\de^2\,h = 0,
\end{equation}
which means that $\de^2\,h = 0$, if $2\,\alpha + \beta\,D \neq 0$. Replacing in the equation \eqref{EomsPartialGaugeFixed}, 
\begin{equation}\label{EomsPartialGaugeFixed2}
-4\,\alpha\,\de^2\,h_{\mu\nu} + (2\,\alpha -\beta)\,\de^\vrho\,\de_{(\mu}\,h_{\nu)\vrho} = 0.
\end{equation}
Imposing the condition \eqref{ConstModelI} on the constants, the previous equation is simply
\begin{equation}
\de^2\,h_{\mu\nu} = 0.
\end{equation}
which is the ghost-number-zero equation in \eqref{EomsPartialGaugeFixed}.

In vector models, the analogue of \eqref{SiegelI}, that is $\mathfrak{b}\,\psi = 0$, on the state $\psi$ in \eqref{psinottracedII} dictates
\begin{equation}
\phi = 0, \quad \tilde{h}_{\mu\nu} = 0, \quad \tilde{\lambda} = 0,
\end{equation}
so that the state in the Siegel gauge is
\begin{equation}
\tilde{\psi} = h_{\mu\nu}\,\alpha^\mu\,\alpha^\nu + \lambda\,B + \tilde{\phi}\,C, \quad\text{such that}\;\mathfrak{b}\,\tilde{\psi} = 0.
\end{equation}
The partially on-shell action for the deformed model is
\begin{align}
S_{\text{Siegel}} &= \int\diff^D x\,\overline{\tilde\psi}\,[\tfrac{1}{2}\,\ell_{\mathrm{w}}(\alpha,\beta) -(2\,\alpha-\beta)\,(C\,\bar{B}-\bar{C}\,B)\,H]\,\tilde\psi \nonumber\\
&= \int\diff^D x\,(-4\,\alpha\,h^{\mu\nu}\,\de^2\,h_{\mu\nu} + 2\,(2\,\alpha - \beta)\,h^{\mu\nu}\,\de^\vrho\,\de_\mu\,h_{\nu\vrho} - 2\,\beta\,\tilde{\phi}\,\de^2\,\lambda).
\end{align}
Again, the undeformed case is obtained by setting $\alpha = -\frac{3}{8}$ and $\beta = \frac{1}{4}$.
The equations of motion are
\begin{equation}
-4\,\alpha\,\de^2\,h_{\mu\nu} + (2\,\alpha - \beta)\,\de^\vrho\,\de_{(\mu}\,h_{\nu)\vrho} = 0,\quad
\de^2\,\tilde{\phi} = 0,\quad
\de^2\,\lambda = 0,
\end{equation}
among which the first two equations, which have ghost number zero, are the same as in \eqref{EomsPartialGaugeFixed2}.

\section{Conclusion and outlook}\label{conclusions}

In this work, we have provided, for the first time, worldline formulations for covariant fracton theories, reproducing their key features from a first-quantized perspective. Using standard BRST quantization techniques, which automatically reproduce the BV spectrum of spacetime fields, we developed two distinct models, the tensor and the vector model, each capable of capturing a specific covariant fracton theory. We then deformed the vector model, adapting techniques originally developed for interacting worldline models, although in the present case the model is free. We have shown that the deformed vector model can reproduce a broader class of covariant fracton theories, exhausting almost all the possible cases. 

We also aimed to illustrate the connection between worldline and spacetime formulations in a self-consistent way, in order to emphasize the potential of this formulation in a non-trivial example. Indeed, as we discussed, the usual procedure for developing a worldline formulation of a field theory required some extensions, in the present case, in order to take into account the free parameters $\alpha$ and $\beta$, which distinguishes the theories in the family of covariant fracton gauge theories \eqref{actionF}.

Thus, the result of the paper suggests that the conventional formulation of quantum field theory in terms of worldline in target space extends well beyond the limits in which the latter was initially conceived, and suggests that the worldline formulation may constitute a potential equivalent formulation of quantum field theory.

To further develop and apply the models we studied, the next natural step involves the construction of the corresponding worldline path integrals. For the tensor model, this is expected to proceed in close analogy with standard worldline models describing spin-$s$ fields, given the similarity in the constraint structure and gauge symmetries. However, it seems not directly feasible to investigate the path integral in configuration space for the vector model (and its deformation), since all classical constraints are quadratic in the momenta and in the bosonic oscillators. This issue requires further investigation. A related issue concerns the possible inclusion of the classical counterpart of the trace constraint in the worldline action. As discussed in the relevant sections, this constraint is essential at the BRST level. However, its inclusion appears to be incompatible with Siegel gauge-fixing. This raises an open question: should the classical action include the trace constraint? Clarifying this point is an important direction for future work.

Regarding future applications, this work may pave the way for efficient uses of the worldline formalism in condensed matter physics, a direction that remains largely unexplored. It is well known that topological field theories play a crucial role in providing effective descriptions of various condensed matter phenomena, especially when defined on manifolds with boundaries \cite{Birmingham:1991ty, Zhang:1992eu, Amoretti:2014iza, Bertolini:2021iku}. Interestingly, similar features appear to be shared by certain non-topological field theories, such as the fracton models considered in this work \cite{Bertolini:2024yur}. While the incorporation of boundaries into the worldline formalism is still under active development \cite{Ahmadiniaz:2022bwy, Manzo:2024gto}, progress in this direction could enable novel and efficient first-quantized approaches to condensed matter effective field theories. A crucial step toward this goal is the construction of suitable worldline formulations for both topological theories, such as the BF model \cite{Cho:2010rk, Bertolini:2022sao}, as well as for non-topological theories, precisely like the fracton ones. The long-term objective may be, for instance, to capture and analyze boundary effects directly within the worldline framework.

Several extensions of the present work could be worth exploring. For example, an interesting aspect which could be explored along the line of this work is to extend further the knowledge of wordline formulation of linearized gravity \cite{Bonezzi:2018box, Bonezzi:2020jjq} by studying a worldline description of the theory invariant only under transverse linearized diffeomorphisms $\delta_{\hat\xi}\,h_{\mu\nu} = \de_\mu\,\hat{\xi}_\nu + \de_\nu\,\hat{\xi}_\mu$, with $\de^\mu\,\hat{\xi}_\mu = 0$. Transverse diffeomorphisms are the orthogonal counterpart of the fracton gauge transformation, and it is the gauge invariance of unimodular gravity, in which the cosmological constant emerges as a constant of integration, whose value is fixed at the boundary, instead of a free parameter in the action.  

Moreover, it would be interesting to study a possible worldline description of covariant fracton theory in the traceless limit, in which the gauge symmetry is given by $\delta_{\Lambda}\,h_{\mu\nu} = \de_\mu\,\de_\nu\,\Lambda$ together with $\delta_{\Sigma}\,h_{\mu\nu} = 2\,\Sigma\,h_{\mu\nu}$. Since the gauge invariance is larger in this case, there are two spacetime ghosts: besides the fracton ghost $\lambda$, there is also a scalar ghost corresponding to the parameter of the Weyl scaling. Interestingly, as is familiar in conformal gravity, the minimal theory does not need a gauge field for the Weyl scaling. So, there are at least two possible ways to proceed in worldline perspective: introducing a single string field, which has in its components both the spacetime ghosts and a single gauge field, or defining a gauge field for the Weyl scaling, and then gauge it away. The last possibility is equivalent to considering the full conformal group as the gauge group of the theory, and then gauge-fix the special conformal transformations. 

Finally, it has already been observed that the fractonic behaviour can be understood within the ``ultralocal” nature of Carrollian physics,\footnote{``Carroll transformations" are the limit of Lorentz transformations by sending the speed of light to zero, \cite{Levy-Leblond:1965dsc}. In Carroll limit, the invariant interval $\diff s^2 = - c^2\,\diff t^2 + \diff x^2$, taking the limit as $c$ tends to zero, becomes $\diff s^2 = \diff x^2$. This means that there exists an absolute space, on which all observers agree. In Carrollian physics, temporal variations dominate over spatial gradients. In the zero-speed limit, light cones are shrunk to the time axis, meaning any causal connection between event points with distinct spatial parts is excluded. In other words, if the speed of light is zero and the only allowed transmissions are those with speed less than that of light, no transmission is allowed, and no event point can be the cause or effect of any event point. For this reason, the Carrollian limit is also called ``ultralocal”.} since fractons are characterized by reduced mobility due to the conservation of the dipole moment \cite{Bidussi:2021nmp}. The possibility of obtaining fractonic Lorentz-breaking models from the covariant formulations has not been explored yet. Now, in \cite{Henneaux:2021yzg} the Carrollian contraction of Lorentz-invariant theories is studied in general: it is shown that there always exist two independent contractions of the same Lorentz-invariant theory, which produce two independent Carrollian theories. To do this, one considers the action of the Lorentz-invariant theory in Hamiltonian formulation, in which Lorentz invariance is not manifest. The two limits consist in turning on only the electric or only the magnetic part of the energy, and are distinguished by how the fields are rescaled with respect to the speed of light. We reserve for the future the analysis of the electric/magnetic Carroll contraction of covariant fracton gauge theories.

\section*{Acknowledgments}

We acknowledge enlightening discussions and helpful insights from Fiorenzo~Bastianelli, Roberto Bonezzi, Olindo~Corradini, Sebastián~A.~Franchino-Vi\~nas, Camillo~Imbimbo, Alessandro~Miccich\`e, and Matteo Vassallo. F.F. is especially grateful to the organizers of the workshop ``New Trends in First Quantization: Field Theory, Gravity and Quantum Computing" (Physikzentrum Bad Honnef) for providing an inspiring atmosphere that played a key role in shaping parts of this research.

\nocite{Frob:2020gdh}

\appendix

\firstsectioneqnums  

\section{Conventions and notations}\label{AppendixConvNot}

We denote graded commutators by
\begin{equation}
[A,B] := A\,B - (-)^{|A||B|}\,B\,A,
\end{equation}
where $|A|$ is the Grassmann parity of the operator $A$, namely $0$ if $A$ is even, and $1$ if $A$ odd. Poisson brackets between phase-space functions $a$ and $b$ are denoted by
\begin{equation}
\{a,b\}:=\frac{\partial a}{\partial x^\mu}\frac{\partial b}{\partial p_\mu}-\frac{\partial a}{\partial p^\mu}\frac{\partial b}{\partial x_\mu}.
\end{equation}
They are extended to take into account ghost variables in the BRST construction as in \cite{VanHolten:2001nj}.

Throughout the paper, we used the same notations for the classical variables and the quantum operators to avoid cluttering, when the distinction is clear from the context. Brackets $(\dots)$ or $[\dots]$ among indices denote symmetrization or antisymmetrization \emph{without} any understood numerical factor.

As regards BV formalism and inner product in graded Hilbert spaces, we follow the conventions in \cite{Barnich:2003wj, Bengtsson:2004cd, Barnich:2004cr, Grigoriev:2006tt, Fuster:2005eg}. We summarize in the following the main features and definitions.

The \emph{BV spectrum} of a field theory whose fields and ghost are $\{\vphi^i\}_i$ is the set $\{\vphi^i,\vphi^*_i\}_i$, where $\vphi^*_i$ is the antifield of $\vphi_i$. We use the following conventions for the \emph{BV bracket} $(\cdot,\cdot)$ and for the \emph{BRST differential operator} $s$:
\begin{equation}
(F,G) = \sum_i\frac{\delta_R F}{\delta\vphi^i}\,\frac{\delta_L G}{\delta\vphi_i^*} - \frac{\delta_R F}{\delta\vphi_i^*}\,\frac{\delta_L G}{\delta\vphi^i},\quad
s = (\Gamma_{\textsc{bv}},\cdot),
\end{equation}
where $F,G$ are any functionals of the fields and antifields, the sum is extended to all the fields of the theory, and $\Gamma_{\textsc{bv}}$ is the \emph{BV action}. The previous definitions imply
\begin{equation}\label{sphisphistar}
s\,\vphi^i = -\frac{\delta_R\Gamma_{\textsc{bv}}}{\delta\vphi_i^*}, \quad
s\,\vphi_i^* = \frac{\delta_R\Gamma_{\textsc{bv}}}{\delta\vphi^i},
\end{equation}
and the BV action can be written as
\begin{equation}
\Gamma_{\textsc{bv}} = \sum_i s\,\vphi^i\,\vphi_i^* + s\,\vphi^*_i\,\vphi^i.
\end{equation}

Let $\psi$ a complex odd variable: $\psi = a + i\,b$, where $a$ and $b$ are real odd variables. The complex conjugate $\psi^*$ requiring $\psi^*\,\psi$ to be real
\begin{equation}
(\psi^*\,\psi)^* = \psi^*\,\psi.
\end{equation}
This implies that 
\begin{equation}
(a\,b)^* = b\,a.
\end{equation}
If $\psi = a+i\,b$ and $\chi = c +i\,d$ are allowed to take matrix values, the complex conjugate, the transpose, and the Hermitian conjugate of their product fulfill
\begin{equation}
(\psi\,\chi)^* = (-)^{\psi\,\chi}\,\psi^*\,\chi^*, \quad
(\psi\,\chi)^t = (-)^{\psi\,\chi}\,\chi^t\,\psi^t, \quad
(\psi\,\chi)^\dagger = \chi^\dagger\,\psi^\dagger.
\end{equation}

Consider a supermanifold with $D$ even coordinates $\lbrace x^\mu\rbrace$ and $n$ odd coordinates $\lbrace\vtheta_i\rbrace$:
\begin{align}
[x^\mu,x^\nu] &= x^\mu\,x^\nu - x^\nu\,x^\mu = 0, \\
[\vtheta_i,\vtheta_j] &= \vtheta_i\,\vtheta_j + \vtheta_j\,\vtheta_i = 0, \\
(-)^{|x^\mu|} &= 1, \quad
(-)^{|\vtheta_i|} = -1.
\end{align}
The right and the left derivatives with respect to a variable $c$ are defined by
\begin{equation}
\frac{\de_L}{\de c}\,(c\,a) = a = \frac{\de_R}{\de c}\,(a\,c).
\end{equation}
The relation between the left and the right derivatives is 
\begin{equation}
\frac{\de_L f}{\de c} = (-)^{|c|(|f|+1)}\,\frac{\de_R f}{\de c},
\end{equation}
which means that, if the variable is even, the two derivatives coincide, as it should be; if the variable is odd, the sign is opposite only in the case in which the derived function is even. If $\overline{c}$ is the complex conjugate of $c$, then
\begin{equation}
\overline{\frac{\de_L}{\de c}} = (-)^{|c|}\,\frac{\de_L}{\de\overline{c}}.
\end{equation}
Since $\frac{\de_L\,c}{\de\,c} = 1$, the left (or equivalently the right) derivative is even/odd if the variable is even/odd
\begin{equation}
\bigg\vert \frac{\de_L}{\de c} \bigg\vert = \bigg\vert \frac{\de_R}{\de c} \bigg\vert = |c|.
\end{equation}
(We assume that the complex conjugation does not change the statistics.) For the same reason, if we assign a ghost number $g(c)$ to the variables $c$, then the left or the right derivative with respect to $c$ has opposite ghost number:
\begin{equation}
g(\tfrac{\de_L}{\de c}) = g(\tfrac{\de_R}{\de c}) = -g(c).
\end{equation}
The Berezin integral with respect to an odd variable $\vtheta$ is defined by
\begin{equation}
\int\diff\vtheta\,1 = 0, \quad
\int\diff\vtheta\,\vtheta = 1,
\end{equation}
that is, it is equivalent to the left derivative. One deduces that
\begin{equation}
\overline{\diff\vtheta} = -\diff\,\overline{\vtheta}, \quad
|\diff\vtheta| = |\vtheta|, \quad
g(\diff\vtheta) = -g(\vtheta).
\end{equation}
Given two wave functions $\Phi = \Phi(x,\vtheta)$ and $\Psi = \Psi(x,\vtheta)$, we define the following inner product 
\begin{equation}\label{InnerProduct}
\langle \Phi,\Psi \rangle = \alpha_n\int\diff\vtheta_1\dots\diff\vtheta_n\,\overline{\Phi}(x,\vtheta)\,\Psi(x,\vtheta),
\end{equation}
where $\alpha_n$ is a constant. Therefore,
\begin{equation}
|\langle\Phi,\Psi\rangle|= n + |\Phi| + |\Psi| \,\text{mod}\,2 \,n. 
\end{equation}
This means that we can assign the following parity to the inner product
\begin{equation}
(-)^{|\langle\cdot,\cdot\rangle|}= (-)^n.
\end{equation}
Similarly, if we assign the same ghost number $g(\vtheta)$ to $\vtheta_i$, for each $i$, then
\begin{equation}
g(\langle\Phi,\Psi\rangle) = -n\,g(\vtheta) + g(\Phi) + g(\Psi), 
\end{equation}
which means that one can assign the following ghost number to the inner product
\begin{equation}
g(\braket{\cdot,\cdot}) = -n\,g(\vtheta).
\end{equation}
Choosing $\alpha_n$ in \eqref{InnerProduct} such that
\begin{equation}\label{AlphaChoice2}
\alpha_{0,3,4,7,8,\dots} = 1,\quad
\alpha_{1,2,5,6,\dots}= i,
\end{equation}
then one can check that the following property holds, if the coordinates are assumed to be real:
\begin{equation}\label{ConjInnerProduct}
\overline{\langle\Phi,\Psi\rangle} = (-)^{n(|\Phi|+|\Psi|)}\,\langle\Psi,\Phi\rangle.
\end{equation}
In particular, in the case $n=1$, choosing $\vtheta_1 = c$, with $|c|=1$, $g(c) = 1$, so that $\alpha_1 = i$, the inner product is
\begin{equation}
\langle\Psi,\Phi\rangle = i\int\diff c\,\overline{\Psi}\,\Phi,
\end{equation}
such that
\begin{equation}
\overline{\langle\Phi,\Psi\rangle} = (-)^{|\Phi|+|\Psi|}\,\langle\Psi,\Phi\rangle.
\end{equation}

\printbibliography

\end{document}